\documentstyle[12pt]{article}
\newcommand{\mysection}[1]{\setcounter{equation}{0}\section{#1}}
\renewcommand{\theequation}{\thesection.\arabic{equation}}
\newcommand{\nc}{\newcommand}
\nc{\beq}{\begin{equation}} \nc{\eeq}{\end{equation}}
\nc{\beqa}{\begin{eqnarray}} \nc{\eeqa}{\end{eqnarray}}
\nc{\lsim}{\begin{array}{c}\,\sim\vspace{-21pt}\\< \end{array}}
\nc{\gsim}{\begin{array}{c}\sim\vspace{-21pt}\\> \end{array}}

\begin{document}

\begin{titlepage}
\begin{center}
{\hbox to\hsize{hep-ph/9501336 \hfill  INFN-FE 01-95}}

{\hbox to\hsize{\hfill MIT-CTP-2404}}

\bigskip

\bigskip

{\Large \bf   Could the Supersymmetric Higgs Particles Naturally be
Pseudo-Goldstone Bosons? \footnotemark[1]      } \\

\bigskip

\bigskip

{\bf Zurab Berezhiani } \\

\smallskip

{\small \it INFN, Sezione di Ferrara, I-44100, Italy,} \\
{\it Institute of Physics, Georgian
Academy of Sciences, 380077 Tbilisi, Georgia}

\smallskip
{\small and}
\bigskip

{\bf Csaba Cs\'aki and  Lisa Randall}\footnotemark[2]\\

\smallskip

{ \small \it Center for Theoretical Physics

Laboratory for Nuclear Science and Department of Physics

Massachusetts Institute of Technology

Cambridge, MA 02139, USA }

 \bigskip

\vspace{2cm}

{\bf Abstract}\\[-0.05in]
\end{center}
The doublet-triplet splitting problem is perhaps the most problematic
aspect of supersymmetric grand unified theories.
It can be argued that the  most natural reason for the  Higgs doublets to
be light is that they are pseudo-Goldstone bosons associated with
the spontaneous breakdown of an accidental global symmetry. In this paper
we discuss the possibility of implementing this idea in the SU(6) model of
 refs. \cite{Zur,Bar2,Bar3,Bar4}.
We show that although it is simple to generate an accidental symmetry of
the  renormalizable terms of the potential, it is quite difficult to construct
a model which allows for the preservation of the accidental symmetry in the
nonrenormalizable terms.
We summarize the constraints on such models and then give three different
ways to construct a superpotential where the dangerous mixing terms are
sufficiently suppressed even in the presence of nonrenormalizable
operators. With these examples we demonstrate the existence of consistent
models implementing the Higgs as pseudo-Goldstone boson scheme. We extend
one of the three examples to include fermion masses.
  We also show that when restricted to  regular group embeddings
the only possible  models without light triplets are trivial generalizations
of the SU(6) model we consider.
\bigskip

\footnotetext[1]{Supported in part
by DOE under cooperative agreement \#DE-FC02-94ER40818.}
\footnotetext[2]{NSF Young Investigator Award, Alfred P.~Sloan
Foundation Fellowship, DOE Outstanding Junior
Investigator Award. }

\end{titlepage}

\renewcommand{\thepage}{\arabic{page}}
\setcounter{page}{1}

\baselineskip=18pt

\mysection{Introduction}

The  unification of the gauge couplings of supersymmetric grand unified
theories at about $10^{16}$ GeV is strong evidence that the standard
model is  embedded in a supersymmetric grand unified theory (GUT).  But
it is only possible to accept the possibility of grand unified theories
if an acceptable reason for the large separation in mass scales between
the Higgs doublet and triplet fields is identified.  Most models which exist
require either  fine tuning or have a very complicated field content and/or
do not take into account the possible nonrenormalizable
operators, suppressed by the Planck mass.

In order to avoid fine tuning, several solutions to the doublet-triplet
splitting problem have been proposed. They include the sliding singlet
\cite{Wit}, the missing partner mechanism \cite{Grin,Ell}, and an SO(10) based
solution \cite{dimwil,Barr}.
The sliding singlet solution has been shown to be unstable
to radiative corrections \cite{Nem}, and is therefore unacceptable.
The missing partner mechanism will in general have problems
when nonrenormalizable operators are included in the Hamiltonian,
which will generate a mass for the doublet Higgs.  The SO(10)
models might work, but
they often require a large field content \cite{Barr} so that asymptotic freedom
of the GUT-theory is destroyed; thus nonrenormalizable operators
are suppressed only by the scale where the GUT-coupling becomes strong
(this scale can be significantly lower than the Planck-scale).

One of the most economical and satisfying explanations for why the Higgs
doublets are light could be that  they are pseudo-Goldstone bosons (PGB's) of a
spontaneously broken accidental global symmetry
of the Higgs sector \cite{Jap}.
The Higgs sector of the chiral superfields is defined with the use of
matter parity. Under this $Z_2$ symmetry all matter fields (fermion
fields) change sign while the Higgs fields are invariant.
When Yukawa couplings are incorporated (couplings of the Higgs sector
to matter fields), the accidental global
symmetry is explicitly broken; however, because of supersymmetric
nonrenormalization theorems the Higgs masses can only be of
order of the supersymmetry breaking, or weak scale.

The first attempts to build such a model were made by requiring that the chiral
 superfields of a given gauge group are put together into a representation of
 a bigger global symmetry
 group \cite{Jap,Ans,Bar1}. For example  the $24, 5, \bar{5}$
 and $1$ of an SU(5) gauge group could form the
$ 35$ adjoint of SU(6). While the
global SU(6) breaks to SU(4)$\otimes$SU(2)$ \otimes$U(1),
the gauged SU(5)
 breaks to SU(3)$\otimes$SU(2)$\otimes$U(1), and the uneaten PGB's are in
two SU(2) doublets \cite{Jap,Ans}.
 Other similar models were discussed in ref. \cite{Bar1}.

Unfortunately this model requires  even more fine tunings of the
parameters of the superpotential than the usual fine tuning
solution of the doublet-triplet splitting problem.
For example in the SU(5) model mentioned above this would mean
that for the general superpotential
\begin{eqnarray}
& & W=\frac{1}{2}M {\rm Tr}\Sigma^2 +\frac{1}{3}\lambda {\rm Tr}
\Sigma^3 +\mu \bar{H} \Sigma H +
\alpha \bar{H}H + \rho_1 Y +\frac{\rho_2}{2}Y^2 + \nonumber \\ & & \rho_3 Y^3 +
 \rho_4 {\rm Tr}\Sigma^2 Y +\rho_5 \bar{H}H Y,
\end{eqnarray}
where the fields $\Sigma ,H,\bar{H}, Y$ are the
SU(5) fields transforming according to
$ 24, 5, \bar{5}, 1$ the following
relations have to hold in order to have the larger global SU(6) invariance:
\begin{eqnarray}
 && \alpha =M=\rho_2,\;\; \mu =\lambda ,\;\; \rho_3=-\frac{2}{3}
 \left( \frac{2}{15}
  \right)^{\frac{1}{2}} \lambda ,\;\; \nonumber \\ &&
\rho_4 =\frac{1}{\sqrt{30}} \lambda ,\;\;   \rho_5 = -2 \left( \frac{2}{15}
 \right)^{\frac{1}{2}} \lambda .
\end{eqnarray}
These relations are very unlikely to be a result of a symmetry
of a higher energy theory. Thus this version does not
tell much more than the original fine tuned SU(5) theory.

A much more appealing scenario is that the accidental symmetry of the
superpotential arises because two sectors of the chiral superfields responsible
for gauge symmetry breaking do not mix and thus the global symmetry of
this sector is $G\otimes G$ instead of the original gauge group $G$
\cite{Zur,Bar2,Bar3,Bar4}.
This accidental symmetry could be a result of a discrete symmetry that forbids
the mixing of the two sectors so this scenario might well be
a consequence of a symmetry of a larger theory.
During spontaneous symmetry breaking $G\otimes G \rightarrow G_1\otimes G_2
$  while the diagonal $G$ (which is the original gauge group) breaks to
SU(3)$\otimes$SU(2)$\otimes$U(1).

The D-terms of the group $G\otimes G$ in this scheme of
spontaneous symmetry breaking (SSB) vanish
in order to preserve supersymmetry. Because supersymmetry is preserved,
the requirement for ``total doubling" \cite{Jap} is fulfilled, so
associated to every Goldstone boson there is
also   a pseudo-Goldstone boson in a chiral multiplet which is massless only
by supersymmetry.
Therefore, all the scalars in a Goldstone  chiral superfield  are light,
not only one of the scalar components. We will refer to both as PGB's
throughout the paper.
The genuine Goldstone bosons remain massless
even after adding the soft SUSY breaking terms, while the
pseudo-Goldstone bosons get masses
at the order of the weak scale at this stage. The remaining massless
states get masses during the running down from the GUT scale to the weak
scale due to the symmetry breaking Yukawa couplings.

 These Yukawa
couplings have to break
the accidental global symmetry of the Higgs sector explicitly. Otherwise
the couplings of the Higgs doublets (which are identified with the uneaten
PGB's of the broken global symmetry) to the light fermions would vanish.
Thus there would be no source for the light fermion masses. The
nonvanishing of the couplings of the Higgs fields to the light fermions
(especially to the top quark) is also essential for radiative electroweak
breaking. Thus it is necessary that in these models the Yukawa couplings
explicitly break the accidental symmetry of the Higgs sector.

Explicit symmetry breaking terms in the Higgs sector can yield additional
contributions to the $\mu$-term of the Higgs potential (the models
presented in section 4 will contain such explicit breaking terms). There
can also be additional contributions to the $\mu$-term from
nonrenormalizable contributions to the K\"ahler potential \cite{Guid}.

An example of models of this kind was given in refs.
\cite{Zur,Bar2,Bar3,Bar4}.
In
this case $G=$ SU(6), and the accidental global symmetry is broken to
SU(4)$\otimes$SU(2)$\otimes $U(1)$\otimes $SU(5). There are  exactly two
light doublets in this model, so  the low energy particle content is just
that of the MSSM. The Higgses are naturally light (they are PGB's),
while the triplets have masses of ${\cal O}(M_{GUT})$.

Although such a model is very appealing in principle,  it
is not clear that it holds up to more detailed scrutiny.
The first problem is to construct a potential with the desired
symmetries and symmetry breakings. The second problem is
to generate a fermion mass spectrum compatible with observation.

The flavor problem can be addressed by enhancing the field
content and including nonrenormalizable operators \cite{Bar3,Bar4}.
However, the first problem is very difficult. The models
of refs. \cite{Bar2,Bar3}  do not give the correct minimum without fine
tuning and are therefore unacceptable.
The model of ref. \cite{Bar4}
has the correct symmetry and gives the desired minimum if
one incorporates only the renormalizable terms in the superpotential.
However, once nonrenormalizable terms are incorporated
it is very difficult to construct acceptable models without fine tuning.

In this paper we outline many  constraints for model building. We consider only
models of the second type; that is, models where the accidental global
symmetry arises as a result of two nonmixing sectors of the Higgs fields.
In supersymmetric theories it may happen that some operators are unexpectedly
missing from the superpotential even though they are allowed by all
symmetries of the theory. However we will use the most pessimistic
assumption, that is, all terms consistent with all the symmetries are
present and they are as big as they can be (suppressed only by
the appropriate powers of $M_{Pl}$).  Our philosophy for constructing models
is to find  discrete symmetries that forbid the dangerous mixings of the
two sectors. These could be either  R-type or usual discrete symmetries.
(We assume that all other symmetries of the theory above the Planck scale
are broken with the exception of some possible discrete symmetries. These
 should actually be gauge type discrete symmetries so that they
are not destroyed by large gravitational corrections. This implies that
these discrete symmetries could possibly have anomalies. However these
discrete anomalies can always be canceled by adding extra gauge singlets
transforming nontrivially under the discrete symmetries.)

 We find
for such models a general feature
that the more one suppresses mixing terms, the more
fine tuning is necessary in the superpotential to maintain the correct
values of the vacuum expectation values (VEV's)
 if the VEV's of all fields in the Higgs sector are comparable.
To find a way out we either need to introduce small mass
parameters or fields with small (or zero) VEV's.

However, the small parameter which can be used to build these
models is necessarily present in models with supersymmetry
breaking.  Based on the analysis of requirements for a successful
model, we show how to exploit the supersymmetry breaking
scale to create models which have an accidental global ${\rm SU}(6)\otimes
{\rm SU}(6)$
symmetry. In these models the Higgs is naturally light, and there
are no problems with the triplets.
Alternatively we will show how to use the second possibility (the
presence of fields with zero VEV's) to build another class of natural
models without the use of any small parameter.

One can ask whether there are perhaps other models that use an
alternative gauge group for
which it is simpler to construct an acceptable potential.  We will
demonstrate that all models that use only regular group embeddings and
have no extra
light triplet fields are trivial
generalizations of the SU(6) model, and are therefore no better (and
probably worse).

The outline of this paper is as follows. In the following section, we review
the SU(6) model and give a simple example for a model that is acceptable
if one incorporates only renormalizable terms into the superpotential.
 In section 3 we first discuss the requirements for
building an acceptable potential and show why it is difficult to
get a natural model. In the first
subsection we consider the use of alternative SU(6) representations to forbid
the dangerous mixing terms while in the second we discuss the possibility
of using restrictive discrete symmetries for this purpose. We draw the
conclusion that if mixing terms are suppressed there must be either
small parameters or fields with zero VEV in the theory.
In section 4 we present three different  models that naturally
fulfill  all the requirements for the
superpotential.

The first model uses a small mass parameter, namely the weak scale, to get
the correct magnitudes of VEV's of the fields in the Higgs sector.
The second model does not use any small mass
parameter, but exploits the presence of fields with zero VEV's to obtain
an acceptable theory. In the third
 model, we assume the appearance of the
GUT scale by an unspecified dynamical origin. With these three models we
demonstrate that the idea of having the Higgses as pseudo-Goldstone
particles can be naturally implemented.
In section 5, we consider the possibilities
for generalizing the SU(6) model to SU(n) groups. Orthogonal groups and
the exceptional group E$_6$ are considered in appendix A. We conclude
in section 6. Appendix B treats the question of generating the fermion
masses.

\mysection{A Review of the  SU(6) Model}

In the SU(6) model of refs. \cite{Zur,Bar2,Bar3,Bar4}, the gauge group of the
high energy GUT theory is SU(6). The accidental symmetry of the Higgs
part of the superpotential arises because there are two sectors involving
two different fields which do not mix in the potential, so that an
accidental global SU(6)$\otimes $SU(6) symmetry is preserved. The fields
suggested in refs. \cite{Zur,Bar2,Bar3} to realize this idea were $\Sigma$ in
an adjoint 35 representation and $H,\bar{H}$ in $6,\bar{6}$ representations
of SU(6). Their SU(5) decomposition is
\begin{eqnarray}
& & \Sigma =  35=  24+6+\bar{6}+1 \nonumber \\
& & H= 6= 5+1 \nonumber \\
& & \bar{H}= \bar{6}= \bar{5}+1 .
\end{eqnarray}

Then one of the sectors consists of the fields $H,\bar{H}$ and the other of
$\Sigma$. The accidental symmetry is realized if mixing terms of the form
$\bar{H}\Sigma H$ are not present in the superpotential. If the fields
$\Sigma$ and $H,\bar{H}$ develop VEV's of the form
\begin{equation}
\label{svev}
\langle \Sigma \rangle = V \left( \begin{array}{cccccc} 1 & & & & & \\
& 1 & & & & \\ & & 1 & & & \\ & & & 1 & & \\ & & & & -2 & \\
& & & & & -2 \end{array} \right), ~~~~
\langle H \rangle =\langle \bar{H} \rangle = U \left( \begin{array}{c}
1\\0\\0\\0\\0\\0 \end{array} \right),
\end{equation}
then one of the global SU(6) factors breaks to
 SU(4)$\otimes $SU(2)$\otimes $U(1),
while the other to SU(5). Together, the VEV's break the gauge group to
SU(3)$\otimes $SU(2)$\otimes $U(1).

The Goldstone bosons (GB's) coming from the breaking
SU(6)$\rightarrow $SU(4)$\otimes $SU(2)$\otimes $U(1)
are (according to their SU(3)$\otimes $SU(2)$\otimes $U(1)
 transformation properties):
\begin{equation}
(\bar{3},2)_{\frac{5}{6}}+(3,2)_{-\frac{5}{6}}+(1,2)_{\frac{1}{2}}+
(1,2)_{-\frac{1}{2}},
\end{equation}
while from the breaking SU(6)$\rightarrow $SU(5) the GB's are
\begin{equation}
(3,1)_{-\frac{1}{3}}+(\bar{3},1)_{\frac{1}{3}}+(1,2)_{\frac{1}{2}}
+(1,2)_{-\frac{1}{2}}+(1,1)_0.
\end{equation}
But the following GB's are eaten by the heavy vector bosons due to the
supersymmetric Higgs mechanism (the gauge symmetry is broken from SU(6) to
SU(3)$\otimes $SU(2)$\otimes $U(1)):
\begin{equation}
(3,1)_{-\frac{1}{3}}+(\bar{3},1)_{\frac{1}{3}}+(3,2)_{-\frac{5}{6}}
+(\bar{3},2)_{\frac{5}{6}}+(1,2)_{\frac{1}{2}}+
(1,2)_{-\frac{1}{2}}+(1,1)_0.
\end{equation}
Thus exactly one pair of doublets remains uneaten which can be identified
with the Higgs fields of the MSSM. One can show that the uneaten doublets
are in the following combinations of the fields $\Sigma ,H,\bar{H}$:

\begin{equation}
\label{higgs}
h_1=\frac{U h_{\Sigma}-3V h_H}{\sqrt{
9V^2+U^2}},
\end{equation}
\begin{equation}
\label{higgs2}
h_2=\frac{U \bar{h}_{\Sigma}-3V
\bar{h}_{\bar{H}}}{\sqrt{
9V^2+U^2}},
\end{equation}
where  $h_H$ and $\bar{h}_{\bar{H}}$ denote
the two doublets living in the SU(6) field $H$ and $\bar{H}$, while
$h_{\Sigma}$ and $\bar{h}_{\Sigma}$ denote
the two doublets living in the SU(6)
adjoint $\Sigma$ .

In order to get the correct order of symmetry breaking we need to have
$\langle \Sigma \rangle \sim M_{GUT}$, $\langle H \rangle = \langle \bar{H}
\rangle > \langle \Sigma \rangle$. In this case the gauge group is broken as

\[ {\rm SU(6)} \rightarrow {\rm SU(5)} \rightarrow {\rm SU(3)}\otimes
{\rm SU(2)} \otimes {\rm U(1)} \]
In the case of opposite ordering of the magnitudes of the VEV's we would get

\[ {\rm SU(6)} \rightarrow {\rm SU(4)}\otimes {\rm SU(2)} \otimes {\rm U(1)}
 \rightarrow {\rm SU(3)}\otimes
{\rm SU(2)}\otimes {\rm U(1)}, \]
which would give unreasonably large threshold correction to the RG values
of $\sin^2 \theta_W$.

The biggest question of this model is how to realize the necessary
suppression of mixing terms like $\bar H \Sigma H$ in the superpotential
and thus achieve the desired vacuum.
We want to find discrete symmetries that forbid the mixing of the
two sectors. These could be either  R-type or usual discrete symmetries.

In ref. \cite{Bar2} a $Z_2$ discrete symmetry $\bar{H}\rightarrow -\bar{H}$,
$S\rightarrow -S$, $H\rightarrow H$, $\Sigma \rightarrow \Sigma$ ($S$ is
an SU(6) singlet) was suggested to forbid the mixing term $\bar{H}\Sigma H$.
But in the supersymmetric limit the $H,\bar{H},S$ VEV's were all zero, so
these VEV's come from the soft breaking terms, and consequently

\begin{equation}
\langle H \rangle \sim (mM_{GUT})^{\frac{1}{2}} \approx 10^8 {\rm GeV},
\end{equation}
where $m$ is a mass parameter of the order of the weak scale. Consequently
unreasonably large fine tuning is needed to obtain
$\langle H \rangle > M_{GUT}$.

One can overcome this problem by introducing more fields into the
theory \cite{Bar4}. One can take for
example two adjoints $\Sigma_1,\Sigma_2$ instead of just one and
a discrete $Z_3$ symmetry under which $\Sigma_1\rightarrow
e^{\frac{2\pi i}{3}}\Sigma_1$ and  $\Sigma_2\rightarrow
e^{-\frac{2\pi i}{3}}\Sigma_2$, while $H,\bar{H},S$ are invariant.
Then the most general renormalizable superpotential is of the form
\begin{eqnarray}
& & W(S,H,\bar{H},\Sigma_1,\Sigma_2)= aS(\bar{H}H-\mu^2)
-\frac{M'}{2} S^2 -\frac{\gamma}{3}S^3-m\bar{H}H +\nonumber \\
& & \alpha S {\rm Tr} \Sigma_1 \Sigma_2 + M{\rm Tr}\Sigma_1\Sigma_2 +
\frac{\lambda_1}{3}{\rm Tr} \Sigma_1^3 +\frac{\lambda_2}{3}{\rm Tr} \Sigma_2^3.
\end{eqnarray}
 which automatically has the global SU(6)$\otimes $SU(6) symmetry.
The VEV's are:
\begin{eqnarray}
& & \langle S \rangle =\frac{m}{a} \\
& & \langle \Sigma_1 \rangle = \frac{\alpha \frac{m}{a}+M}{(\lambda_1^2
\lambda_2)^{\frac{1}{3}}} \left( \begin{array}{cccccc} 1& & & & & \\
 &1 & & & & \\ & & 1& & & \\ & & & 1& & \\ & & & & -2 & \\ & & & & & -2
\end{array} \right) \\
& & \langle \Sigma_2 \rangle = \frac{\alpha \frac{m}{a}+M}{(\lambda_2^2
\lambda_1)^{\frac{1}{3}}} \left( \begin{array}{cccccc} 1& & & & & \\
 &1 & & & & \\ & & 1& & & \\ & & & 1& & \\ & & & & -2 & \\ & & & & & -2
\end{array} \right) \\
& & \langle H \rangle = \langle \bar{H} \rangle = \left[
\mu^2 +\frac{M'm}{a^2} +\frac{\gamma m^2}{a^3} - \frac{12(\alpha \frac{m}{a}+
M)^2\alpha m}{a^2 \lambda_1 \lambda_2} \right]^{\frac{1}{2}}
 \left( \begin{array}{c} 1 \\ 0 \\
0 \\ 0 \\ 0 \\ 0 \end{array} \right).
\end{eqnarray}
so that this model gives the correct order of VEV's if $m,M,M' \sim
M_{GUT}$. There is no renormalizable mixing term allowed by the
discrete symmetry that could destroy the accidental SU(6)$\otimes$SU(6)
symmetry.

But the problem is that it is not sufficient to consider only
renormalizable operators. Nonrenormalizable operators scaled by inverse
power of $M_{Pl}$
can potentially introduce large breaking of the global SU(6)$\otimes$SU(6)
 symmetry, which in turn yields large contributions to the PGB masses.
In particular, in the above example the term
$\frac{1}{M_{Pl}} \bar{H} \Sigma_1\Sigma_2 H$ is allowed and gives an
unacceptably big correction to the PGB masses if present. Namely, the
Higgs doublets would acquire masses $\sim M_{GUT}^2/M_{Pl}
\approx 10^{13}~{\rm GeV}$.

\section{Requirements and Constraints for the Superpotential}

We have seen in the previous section that even if
mixing terms of the renormalizable superpotential
are forbidden by some discrete
symmetries, the possible nonrenormalizable operators can still break
the accidental global symmetry and thus spoil the solution to the doublet
triplet splitting problem.

The origin of the nonrenormalizable operators can be of two forms: they
either come from integrating out heavy (${\cal O}(M_{Pl})$) particles from
tree level diagrams or they can be a consequence of nonperturbative effects.

The dangerous mixing terms coming from integrating out the heavy fields
can be easily forbidden  by some additional requirements on
the Planck scale particles, for example by requiring
that all the Planck mass fields are matter (fermion) fields.
In this case the nonrenormalizable terms arising from integrating out the
heavy fields  can only yield Yukawa terms. But we know that Yukawa terms are
irrelevant from the point of view of PGB masses (the accidental global symmetry
is a symmetry of the Higgs sector only). This assumption on the
heavy fields is usually fulfilled by the interaction terms introduced in
models for light fermion masses (e.g. \cite{Bar3,Bar4}). In those models
we want to generate exactly additional Yukawa terms suppressed by Planck
masses. Thus matter parity can be used to forbid all dangerous
nonrenormalizable mixing terms
arising from tree diagrams. Loop diagrams are naturally proportional to
supersymmetry breaking.

Even if the above assumption for the superpotential involving heavy
fields is valid, there is still the possibility of Planck
mass suppressed operators in the superpotential which violate
the global symmetry. Although the nonrenormalization theorem
prevents these operators from being generated perturbatively
if they were not present at tree level, we will take the attitude
that all operators consistent with the low energy gauge and
discrete symmetries are present, both in the K\"ahler potential
and in the superpotential.  We ask the question whether it
is possible with this assumption to still maintain an approximate
global symmetry which can guarantee that the Higgs doublet is
sufficiently light.

The first observation is that the K\"ahler potential will always
permit symmetry breaking terms, suppressed only by two powers
of $M_{Pl}$, for example:
\beq
{1 \over M_{Pl}^2} H^\dagger \Sigma^\dagger \Sigma H .
\eeq
There is no symmetry which can prevent such a term. However, although
such terms  do break the accidental global symmetry, they
do not lead to generation of a mass term for the PGB's.

However, the PGB mass terms will be generated if the global symmetry
is broken in the superpotential.
In the remainder of this section, we show that it is extremely
difficult to prevent mixing in the superpotential.

We now summarize the requirements for the superpotential of a realistic model.

1. The mass terms for the PGB's (which are identified with the Higgs doublets
of the MSSM) resulting from the symmetry breaking mixing terms should be
suppressed at least by a factor of $10^{-13}$ compared to the GUT scale.
 In this case the masses of the PGB's will be at the order of 1000 GeV.

2. The VEV's of the fields $\Sigma$ and $\bar{H},H$ should be naturally
(without tuning) at the order of the GUT scale ($10^{16}$ GeV).

3. The triplets contained in the $\Sigma$ field should have GUT-scale
masses not to cause too large proton decay. One might think that the same
requirement holds for the triplets contained in the fields $\bar{H},H$.
 However these triplets are eaten by the heavy SU(6) gauge bosons and are not
dangerous for proton decay.

There are two  approaches one could imagine to prevent
mixing through nonrenormalizable operators. One might
try to find a representation of
SU(6) which   breaks  SU(6) to
SU(4)$\otimes $SU(2)$\otimes $U(1) but does not  allow mixing.
Alternatively,  one can search for
more restrictive discrete symmetries.

  We have
found no solution with alternative representations.
It is also very difficult to realize the second solution if we try to use
$M_{Pl}$ as the only mass scale in the theory.
We show that one either needs to introduce small mass scales into the
theory or to use fields that have zero VEV's to overcome all constraints
listed in the following subsections.

In the next two subsections we consider the above two possibilities for
model building. We show that the above requirements
necessarily lead us to consider the kind of models
 presented  in section 4.

\subsection{Alternative representations}

Let us first consider the possibility of achieving the desired
symmetry breaking pattern with alternative representations of SU(6).
One can consider symmetric, antisymmetric, or mixed representations.
We don't want to replace the the $H,\bar{H}$
fields because the $\bar{H}$ field is capable of splitting the light
fermions from the heavy ones  through the renormalizable operator
$15\bar{H}\bar{6}$, see \cite{Bar3}, so we only consider
replacing the $\Sigma$ field.
If the representation is symmetric, one does not achieve the desired
symmetry breaking pattern.  An antisymmetric representation (for
example a 15 looks promising)  can achieve a good symmetry
breaking pattern  since it can break SU(6) to SU(4)$\otimes $SU(2).
 Thus with an additional
U(1) gauge group SU(6)$\otimes $U(1) could break
to SU(4)$\otimes $SU(2)$\otimes $U(1)
(much like the flipped SU(5) model of \cite{Ell}).
Furthermore, it looks naively as if it can
forbid undesired mixing terms such as $15 \bar{6} \bar{6}$  because of
the antisymmetry of $15_{ij}$ .
However,  in order to cancel anomalies,
one must introduce additional fields, either $\bar{15}$ or $\bar{6}
+\bar{6}'$. But this addition makes mixing already possible through
$15\; \bar{15}\bar{H}H$ or in the other case through $15 \bar{H} \bar{6}$.
Larger representations do not help because we require a representation
that is capable to break SU(6) to SU(4)$\otimes$SU(2)$\otimes$U(1).

\subsection{Discrete symmetries}

The next possibility is to look for more restrictive discrete symmetries.
Throughout this subsection we will assume that the only mass scale present
in the theory is $M_{Pl}$ and that all fields have VEV's of the order of
the GUT scale.
It turns out that under these assumptions
even with additional fields, it is extremely difficult
to find a satisfactory superpotential with no unnaturally small parameter.
 We first summarize the
reasons why it is difficult to find a satisfactory potential
without fine tuning.  We subsequently elaborate and illustrate
each point in more detail. To be explicit, we assume all fields in the Higgs
sector
have VEV's of order  $10^{-3}M_{Pl}$, the lowest possible value,
in order to obtain the maximum suppression in higher dimension
mixing operators. This ratio might in fact be larger; one
would then need to suppress mixing operators still further.  For
this value, we require that the mixing term is at least of dimension
four greater than the terms in the superpotential which respect
the symmetry and generate the VEV's for the $\Sigma$ and $H, \bar{H}$
fields. The PGB masses will then be at most $(M_{GUT}/M_{Pl})^4 M_{GUT}
\sim 1000$ GeV.

 It is
easy to see that with just the fields $\Sigma, H,\bar{H}$ we can not obtain
a successful superpotential. The reason for this is that in order to get
nonzero VEV's for the fields we need to have at least two terms in both
sectors of the accidental global symmetry (one sector contains the adjoint
$\Sigma$ and breaks SU(6) to SU(4)$\otimes$SU(2)$\otimes$U(1) while
the other $H$ and $\bar{H}$ and breaks SU(6) to SU(5)).
Then the quotient of the two terms can always multiply a term in the
other sector, thereby generating unwanted mixing. Explicitly,
if there are  terms in one sector of the
form
\begin{equation}
\frac{1}{M_{Pl}^{a-3}} {\rm Tr} \Sigma^a + \frac{1}{M_{Pl}^{b-3}}
 {\rm Tr} \Sigma^b,\label{bal}
\end{equation}
then ${\rm Tr} \Sigma^{b-a}$  transforms  trivially under an abelian discrete
symmetry (even if it is an R-type symmetry). The presence of the terms
\begin{equation}
\frac{1}{M_{Pl}^{2c-3}}(\bar{H}H)^c+\frac{1}{M_{Pl}^{2d-3}} (\bar{H}H)^d
\end{equation}
in the other sector then means that terms such as
\begin{equation}
\frac{1}{M_{Pl}^{2c+b-a-3}} (\bar{H}\Sigma^{b-a}H) (\bar{H}H)^{c-1}
\end{equation}
are allowed. The number $b-a$ cannot be arbitrarily big if the dimensionful
fields have VEV's of order $M_{GUT}$.  This is because
in order to balance the two terms in equation
\ref{bal}, there must be a small coefficient of order $\epsilon_G^{b-a}$
where $\epsilon_G=M_{GUT}/M_{Pl}\approx 10^{-3}$.
So in order for the  mixing term to be  suppressed by $10^{-13}$,
the mixing must be suppressed by at least $\epsilon_G^4$.
But then $b-a\ge 4$ and there must be a small parameter in the potential
of order $\epsilon_G ^4$, which is badly fine tuned.

So we  have  established that one requires additional fields,
that there must be at least two operators in each of the two nonmixing sectors
(one involving the $H$ and $\bar{H}$ fields and one involving only the
$\Sigma$ field), and that the quotient of operators in the  superpotential
from the same sector must involve negative powers  of at least one field,
 so that such symmetry invariants are not  holomorphic functions of the fields.

The next point is that in order to prevent fine tuning,
the superpotential should contain operators of
similar dimension.  The argument which we just gave without
additional singlets can readily be generalized (if the singlet VEV
is of the same order as those of other fields) to show that
 in order to prevent fine tuning, the dimension of the operators
in the potential which are balanced at the minimum should have
comparable dimension.   Furthermore, a term of dimension
$d$ will yield a mass term for the
non PGB's of the order $M \approx M_{GUT}\epsilon_G^{d-3}$.   A very
high dimension operator without a large coefficient will yield
masses for the triplet fields much less than $M_{GUT}$. Thus according to
our requirement 3 the terms containing the $\Sigma$ field should have low
dimensions so that the triplets contained in $\Sigma$ have sufficiently
large masses.

Of course, one can consider cases where not all VEV's are the
same, but then VEV's are larger than $M_{GUT}$ and mixing
terms will be less suppressed.

We can generalize the above argument about the superpotential containing
only the fields $\Sigma ,\bar{H},H$ to the case when the superpotential
also includes an additional SU(6) singlet.
To have nonzero VEV's for the fields we
need at least two terms that contain $\bar{H}H$ and two that contain
$\Sigma $ in the superpotential, while all these four terms may contain
the SU(6) singlet field $S$. Thus generally the superpotential will have
the form (if there is no mixing of the two sectors)
\begin{equation}
\label{suppot}
(\bar{H}H)^aS^b+(\bar{H}H)^cS^d+{\rm Tr}\Sigma^eS^f+{\rm Tr} \Sigma^gS^h .
\end{equation}

Without loss of generality we can assume that $d>b$, $f>h$ and $d>f$.
If $f>h$ we require that $g>e$; otherwise the operator ${\rm Tr}
\Sigma^{e-g}S^{f-h}$
(which is just the quotient of the last two terms and thus invariant under
all discrete symmetries) would be holomorphic and could multiply either term
of the $\Sigma$ sector to give a nonsuppressed mixing term. But because $g>e$,
the operator  $\Sigma^{g-e}S^{h-f+d}(\bar{H}H)^c$ is holomorphic. This
operator is
allowed by the discrete symmetries, because it is
the product of two terms of the
superpotential divided by a third term. Therefore the dimension of the
allowed mixing term is equal to the dimension of one of the terms
originally present in the superpotential plus the difference of the
dimension of two terms present in the superpotential. (It is easy to see
that this is also true for the case $e>g$.)
 Thus the necessary fine tuning is equal to
the suppression factor of the mixing term. If we want to suppress mixing
by $\epsilon_G^4$ we will need fine tuning of the same order (to balance
terms of different dimensions). To illustrate
this argument we present a model where although mixing terms are suppressed
 sufficiently  we need unreasonably large fine tuning to get the
 correct VEV's.
In this model the superpotential is given by
\begin{eqnarray}
\label{model}
& & W(\Sigma ,H,\bar{H},S)=\frac{\alpha}{M_{Pl}^4} S^5 \bar{H}H +
\beta S {\rm Tr}\Sigma^2 +\nonumber \\ & & \frac{\gamma}{M_{Pl}^3}
(\bar{H}H)^3 +
\frac{\delta}{M_{Pl}^4} {\rm Tr}\Sigma^7 ,
\end{eqnarray}
where the discrete charges for the fields $\Sigma , \bar{H}H, S$ are
$Q_{\Sigma}=\frac{25}{61}, Q_{\bar{H}H}=\frac{38}{61},Q_{S}=\frac{3}{61}$
and the R-charge of the superpotential is $\frac{53}{61}$.
(The transformation of the fields under the discrete symmetry is
given by $\Phi \rightarrow e^{2\pi iQ}\Phi$.) Then the
first allowed mixing term is $\bar{H}\Sigma^5HS^4$, suppressed by 4 dimensions
compared to ${\rm Tr}\Sigma^7$ or $S^5\bar{H}H$. The equations of motion for
this theory are
\begin{eqnarray}
& & \frac{5\alpha S^4 (\bar{H}H)}{M_{Pl}^4} + \beta {\rm Tr} \Sigma^2=0
\nonumber \\
& & \frac{\alpha S^5}{M_{Pl}} +3\gamma (\bar{H}H)^2=0 \nonumber \\
& & 2\beta S \Sigma +\frac{7\delta}{M_{Pl}^4} (\Sigma^6-\frac{1}{6}
{\rm Tr} \Sigma^6)=0.
\end{eqnarray}
If the VEV of $\Sigma $ has the form

\begin{equation}
\langle \Sigma \rangle = V \left( \begin{array}{cccccc} 1 & & & & & \\
& 1 & & & & \\ & & 1 & & & \\ & & & 1 & & \\ & & & & -2 & \\ & & & & & -2
\end{array} \right)
\end{equation}
then the solution for $V$ is
\begin{equation}
\label{tunevev}
V=M_{Pl} \left[ \left( \frac{-3\gamma}{\alpha} \right) \left(
\frac{12\beta}{5\alpha}\right)^2 \left( \frac{2\beta}{147\delta} \right)^{13}
\right]^{\frac{1}{61}}.
\end{equation}
The number multiplying $M_{Pl}$ should be $10^{-3}$, so even if we assume that
this is the 13/61st power of a combination of the parameters this combination
must be $(10^{-3})^{\frac{61}{13}} \sim 10^{-12}$. Thus we can see
explicitly in this model that the amount of fine tuning ($10^{-12}$) is
equal to the suppression factor of the dangerous mixing terms.

One might think that we can overcome this problem by introducing even more
fields into the theory. If we could find a superpotential where the number
of terms contained in the superpotential is equal to the number of fields
in the superpotential we could assign arbitrarily different R-charges
to the fields in the superpotential and thus forbid mixing terms. However
this is not possible. The reason is the following: suppose we have n fields
and n polynomial terms in the superpotential. Let's call these terms $A_i,
i=1,\ldots {\rm n}$, where $A_i$ is a polynomial of the fields $\Phi_k,
k=1,\ldots {\rm n}$. The superpotential is then
\begin{equation}
\label{nterms}
W(\Phi_i)=\sum_{k=1}^n \alpha_k A_k(\Phi_i).
\end{equation}
The equations of motion are $\frac{\partial W}{\partial \Phi_i}=0$. If
neither of the VEV's is zero then  we can
also write these equations
 in the form $\Phi_i \frac{\partial W}{\partial \Phi_i}=0,
i=1,\ldots {\rm n}$. Thus we get a system of equations
\begin{equation}
\label{sys}
\sum_{k=1}^n \beta_{ik} \bar{A_k} =0,~~ i=1,\ldots {\rm n},
\end{equation}
where $\bar{A_k}={A_k}(\langle \Phi_j \rangle )$.
This is a set of  n linear homogeneous equations for the terms $\bar{A_k}$.
There are two possibilities: the determinant of
the coefficients $\beta_{ik}$ is
either zero or nonzero. To have it zero requires fine tuning of the
parameters in the superpotential and even then we can not have all VEV's
determined by the superpotential because the equations are linearly
dependent so there are in fact fewer equations than n.
If the determinant is nonzero then the only possibility is to have
$\bar{A_k}=0$ for $k=1,\ldots {\rm n}$. This implies
that at least one of the VEV's is zero contrary to our assumption.

Thus we need at least n+1 terms in the superpotential to have the VEV's of
all fields determined of the correct size without fine tuning.
But this means that we can not choose the R-charges
of the fields arbitrarily. Generally these connections among the
R-charges make it very difficult to find an acceptable superpotential that
both determines the VEV's at the right scale without fine tuning and has
the mixing terms sufficiently suppressed. In all cases we examined
with only low dimensional operators for the $\Sigma$ field in the
superpotential
we were either able to find allowed
unsuppressed mixing terms or fine tuning was required to set the VEV's to the
right scale.

\section{Three Models in which the Supersymmetric Higgs Particles are
Naturally Pseudo-Goldstone Bosons}

We have shown in the preceding section that one cannot
construct a model based on low energy discrete symmetries
without  a small parameter if neither of the VEV's of the fields of the
Higgs sector is zero.
However, low energy supersymmetry {\it must} contain a small
parameter, namely the weak scale, or equivalently, the supersymmetry
breaking scale.  In the first subsection, we show how one can exploit
this small parameter to generate models which naturally
respect the accidental  global SU(6)$\otimes$SU(6) symmetry.
Our models differ from the model in ref. \cite{Bar2} in that we
exploit the supersymmetry breaking scale, but we do {\it not}
need to tune the parameters. We naturally balance small terms
against each other.

In the second subsection we present a different class of models. These
models contain fields with zero VEV's; thus the no-go arguments of the
previous section are not valid here. These models include two mass
parameters: all mass terms are proportional to the GUT-scale while the
nonrenormalizable operators are suppressed by the Planck-scale.
 In the
third model  we assume the appearance of a dynamical scale (related to the
GUT scale) but do not specify its origin.

These three  models serve as existence proofs for models
which implement the SU(6)$\otimes$ SU(6) symmetry.
    Based on the considerations of the previous section,
we expect the simplest successful models will have features of one of
 the  models  presented below.

\subsection{Model 1}

In these models we give superpotentials which together with the soft
breaking terms give the correct values of VEV's. This is similar to the
model of ref. \cite{Bar2} but there the superpotential contained only
renormalizable terms. Consequently  the soft breaking terms alone were not
enough to set the VEV's to the right scale and
additional fine tuning was required.

 The essential observation is that
the triplets from $H,\bar{H}$ are eaten by the heavy gauge bosons and
thus we don't need ${\cal O}(M_{GUT})$ mass terms for these fields.
The  superpotential is given by
\begin{equation}
\label{softpot}
W(\Sigma, \bar{H},H)=\frac{1}{2}M{\rm Tr} \Sigma^2+\frac{1}{3}\lambda{\rm Tr}
 \Sigma^3 +\alpha\frac{(\bar{H}H)^n}{M_{Pl}^{2n-3}}
\end{equation}
If one assigns a discrete $Z_n$ symmetry under which  $\bar{H}H \rightarrow
e^{2\pi i/n} \bar{H}H $ and $\Sigma$ is invariant  then
these terms are the lowest order allowed ones. In the supersymmetric limit
\begin{eqnarray}
\langle\Sigma\rangle=\frac{M}{\lambda}\left(\begin{array}{cccccc} 1 & & &
& & \\ & 1 & & & & \\ & & 1 & & & \\ & & & 1 & & \\ & & & & -2 & \\
& & & & & -2 \end{array}\right), ~~~~
 \langle H\rangle =\langle\bar{H}\rangle=0.
\end{eqnarray}
The scalar potential (including the soft breaking terms) will have the form:
\begin{eqnarray}
\label{softbr}
& & V(\Sigma ,\bar{H},H)={\rm Tr}|M\Sigma +\lambda\Sigma^2-\frac{1}{6}\lambda
{\rm Tr}\Sigma^2|^2
+\frac{n^2\alpha^2}{M_{Pl}^{4n-6}} (\bar{H}H)^{2n-2} (|H|^2+|\bar{H}|^2)+
\nonumber
\\ & & A m \lambda {\rm Tr}\Sigma^3+ A'm \alpha
\frac{(\bar{H}H)^n}{M_{Pl}^{2n-3}}+ BMm\Sigma^2+
m^2({\rm Tr}\Sigma^2 +|H|^2+|\bar{H}|^2) +{\rm D-terms} \nonumber \\
\end{eqnarray}
where $m$ is a mass parameter of the order of the weak scale while $A,A',B$ are
dimensionless parameters. The D-terms have to vanish not to have
supersymmetry breaking in the visible sector.
 The soft breaking terms shift the $\Sigma$ VEV only
by a small ($\sim m$) amount.
 However for the $\bar{H},H$ terms we have the
possibility of a new minimum appearing due to the soft breaking terms. To find
this we minimize the $\bar{H},H$ part of the potential
(using $\langle H\rangle=
\langle\bar{H}\rangle = U (1,0,0,0,0,0)$ which is a consequence of the
vanishing of the D-terms).
\begin{equation}
V(U)=\frac{2n^2|\alpha |^2}{M_{Pl}^{4n-6}}U^{4n-2} +
\frac{A'm\alpha}{M_{Pl}^{2n-3}} U^{2n} +2m^2 U^2.
\end{equation}
Minimizing this potential we will get for one of the minima

\begin{equation}
\label{Wval}
U=\left[ \frac{1}{2n^2\alpha (4n-2)} (-nA'+\sqrt{n^2
(A')^2-16n^2(2n-1)}) M_{Pl}^{2n-3}\mu \right]^{\frac{1}{2n-2}}
\end{equation}
The magnitude of the VEV is determined by the factor

\begin{equation} \left[ M_{Pl}^{2n-3}\mu \right]^{\frac{1}{2n-2}}=
\left(\frac{\mu}{M_{Pl}}\right)^{\frac{1}{2n-2}}M_{Pl}
\end{equation}
For $n<4$ we get a smaller scale than $M_{GUT}$ which is not acceptable.
However for $n\geq 4$ the resulting scale always lies between the GUT
scale and the Planck scale. (For $n=4,5,6$ we get $U\approx 1.5~10^{16},
7~10^{16}, 2~10^{17}$ GeV.) Thus all these cases yield naturally the
correct values of the $H,\bar{H}$ VEV's.
The first mixing term allowed by the $Z_n$ symmetry is
$\frac{1}{M_{Pl}^{2n-2}} (\bar{H}H)^{n-1} (\bar{H}\Sigma H)$. However the
resulting mass for the PGB's is
\begin{equation}
U\left( \frac{U}{M_{Pl}} \right)^{2n-3} = \mu \left( \frac{\mu}{M_{Pl}}
\right)^{2n-2} < \mu.
\end{equation}
This means that all models with $n\geq 4$ yield an acceptable theory with
the correct order of VEV's and naturally suppressed mixing terms. The
possibility that the $H,\bar{H}$ VEV's are between the GUT and the
Planck scale may even be welcome from the point of view of fermion masses
(see ref. \cite{Bar3}), and $\langle H \rangle > M_{GUT}$ is also
required for the unification of couplings.

We can not use this method for getting GUT-scale VEV's for the sector
containing the field $\Sigma$ because then the triplets of $\Sigma$
would get too small masses and would spoil the proton stability.
This is however not the case
for the $\bar{H},H$ fields because the triplets from $\bar{H},H$ are
eaten by the SU(6) gauge bosons. Fortunately it suffices to use this
method for only one of the sectors because then mixing terms are already
sufficiently suppressed. This leads us to the choice of the operator
$(\bar{H}H)^n/M_{Pl}^{2n-3}$, while we have no restriction for the other
sector. Alternatively we could use for example the superpotential

\begin{equation}
\label{goodsup}
M{\rm Tr}\Sigma_1\Sigma_2 +\frac{1}{3}\lambda_1{\rm Tr}\Sigma_1^3
+\frac{1}{3} \lambda_2  {\rm Tr}\Sigma_2^3 +\frac{\alpha}{M_{Pl}^{2n-3}}
(\bar{H}H)^n
\end{equation}
with the additional $Z_3$ discrete  charges $Q_{\Sigma_1}=1/3$,
$Q_{\Sigma_2}=-1/3$, $Q_{\bar{H}H}=0$
(similar to the model of ref. \cite{Bar4}).
In this case the mixing term is even more suppressed by the additional
discrete symmetry. The lowest order mixing term in this case is
\[ (\bar{H}\Sigma_1\Sigma_2H) (\bar{H}H)^{n-1}/M_{Pl}^{2n-1}.\]
This model will be used in Appendix B, when we extend it to incorporate
fermion masses.

\subsection{Model 2}

In this class of models we will use low dimension operators to get the
VEV's of the adjoint sector and then use two singlets with zero VEV's to
communicate the required values of the VEV's to the $H,\bar{H}$ fields.
The adjoint sector consists of two adjoint fields $\Sigma_1, \Sigma_2$
and two SU(6) singlets $A,B$, while we introduce additional singlets
($N,T,S$) to get the desired VEV's  for  $H,\bar{H}$. We use a
$Z_3^{(1)}\otimes Z_3^{(2)}\otimes Z_2 \otimes R$ symmetry, where $R$
is a discrete R-symmetry with the charge of the superpotential being
$\frac{1}{3}$. The discrete charge assignments of the fields are given
in table 1.

\begin{table}
\label{charges}
\[ \begin{array}{|c|c|c|c|c|c|c|c|c|} \hline & \Sigma_1 & \Sigma_2 & A &
B & T & S & N & \bar{H}H \\ \hline  Z_3^{(1)} & \frac{1}{3} & -\frac{1}{3}
& -\frac{1}{3} & \frac{1}{3} & 0 & 0 & 0 & 0 \\ \hline  Z_3^{(2)} & 0 & 0
& 0 & 0 & \frac{1}{3} & -\frac{1}{3} & 0 & \frac{1}{6} \\ \hline  Z_2 &
0 & 0 & 0 & 0 & 0 & 0 & 0 & \frac{1}{2}  \\ \hline  R & \frac{1}{9} &
\frac{2}{9} & -\frac{1}{9} & \frac{4}{9} & \frac{5}{9} & \frac{1}{9} &
\frac{2}{3} & \frac{1}{9} \\ \hline  \end{array} \]

\caption{ The discrete charge assignments of the fields of the Higgs
sector of model 2.}
\end{table}
The lowest order superpotential allowed by the discrete and gauge
symmetries is
\begin{eqnarray}
& & M {\rm Tr} \Sigma_1\Sigma_2 +a{\rm Tr}\Sigma_1^3 +
b{\rm Tr}\Sigma_2^2 A + M'A B +c B^3 +\nonumber \\ & &
\frac{\alpha}{M_{Pl}} N ({\rm Tr} \Sigma_2^3 +\beta T^3)
+\frac{\gamma}{M_{Pl}^2} S (T^4-\delta (\bar{H}H)^2).
\end{eqnarray}
The VEV's are
\begin{eqnarray}
& & V_1 = \left( \frac{M^6M'^3}{3^82^5a^5b^3c} \right)^{\frac{1}{9}}
\approx [M^6M'^3]^{\frac{1}{9}} \nonumber \\
& & V_2=\frac{3a}{M}V_1^2 \approx [M^3M'^6]^{\frac{1}{9}} \nonumber \\
& & \langle B\rangle
= -\frac{108 ba^2}{M^2M'}V_1^4 \approx [M^6M'^3]^{\frac{1}{9}} \nonumber \\
& & \langle A\rangle  = -\frac{3c}{M'} \langle B\rangle^2
\approx [M^3M'^6]^{\frac{1}{9}}\nonumber \\
& & \langle T \rangle = (\frac{12}{\beta})^{\frac{1}{3}} V_2 \approx
[M^3M'^6]^{\frac{1}{9}} \nonumber \\
& & \langle H \rangle = \langle \bar{H} \rangle = \frac{\langle T \rangle
}{\delta^{\frac{1}{4}}} \approx [M^3 M'^6]^{\frac{1}{9}} \nonumber \\
& & \langle S \rangle = \langle N \rangle = 0
\end{eqnarray}
where $V_1$ and $V_2$ are defined by
\begin{equation}
\langle\Sigma_1\rangle=V_1\left(\begin{array}{cccccc} 1 & & &
& & \\ & 1 & & & & \\ & & 1 & & & \\ & & & 1 & & \\ & & & & -2 & \\
& & & & & -2 \end{array}\right),~~~~\langle\Sigma_2\rangle=V_2
\left(\begin{array}{cccccc} 1 & & &
& & \\ & 1 & & & & \\ & & 1 & & & \\ & & & 1 & & \\ & & & & -2 & \\
& & & & & -2 \end{array}\right)
\end{equation}
If $M,M' \approx M_{GUT}$ then all fields (with the exception of $N$ and
$S$) have ${\cal O} (M_{GUT})$ VEV's. The lowest possible mixing term in
the superpotential is
$ (\bar{H}\Sigma_1 H)(\bar{H}H) A S$ which yields a supersymmetric mass
term (so called $\mu$-term) for PGB Higgs doublets
$\mu \sim (10^{16}/2~10^{19})^4 ~10^{16} \approx 1000$ GeV. One can see
that the dangerous mixing term is quite big (compared to the
lowest order mixing term of Model 1). One might need some
additional suppression  factor but no large fine tuning.
The feature of this model that there are symmetry breaking terms that yield
extra $\mu$-terms for the Higgs doublets
(which may also arise in the models presented in the previous
subsection) solve a potential problem of these models. Namely, if
there are no explicit symmetry breaking terms in the Higgs sector then
the 'genuine GB's' will remain exactly massless at the GUT scale even
after adding the soft breaking terms.
This results in a potential instability of the Higgs potential
(a flat direction for $h_1=h_2^*$), which has to be removed by radiative
corrections (essentially due to the large top Yukawa coupling).
Explicit global symmetry breaking terms in these models lift this flat
direction and remove the instability. However,
these symmetry breaking terms at the same time invalidate the specific
prediction of the 'Higgs as PGB' scheme (the $\mu$-term is not related
to soft SUSY breaking mass term anymore), and we will
be left with the general Higgs potential of the MSSM.

In the above model all fields (except $S$ and $N$) had the same order of
VEV's, thus there is no hierarchy between the $H, \bar{H}$ and $\Sigma$
VEV's. However such a hierarchy
may be an attractive feature for generating fermion masses and is also
necessary for the unification of couplings. This can be easily achieved
in this model by modifying the discrete charges of $\bar{H}H$. We take
the $Z_3^{(1)}\otimes Z_3^{(2)}\otimes Z_4 \otimes R$ charges for the
$\bar{H}H$ as $0,\; \frac{1}{12}, \; \frac{1}{4},\; \frac{1}{18}$ instead
of the charges listed in table 1 (and all other charges are
unchanged). Then the only change will be that instead of $S(\bar{H}H)^2$
we have $S(\bar{H}H)^4$ appearing in the superpotential. This will result
in an $H$ VEV that is the geometric mean value of $M_{Pl}$ and $M_{GUT}$,
which is desirable for fermion masses. The mixing terms again  yield
${\cal O}(1000 ~{\rm GeV})$ PGB masses.

\subsection{Model 3}

In the third model we assume that some SU(6) singlet
fields have VEV's of the order of the GUT scale through
some unspecified dynamics.

One   possibility to suppress mixing terms is to have at least two fields
whose VEV's  are naturally zero in the supersymmetric limit and whose
presence is required in all dangerous mixing terms. In this case
the mixing terms have the form $S T (\bar{H}H)^a \Sigma^b$ where $S,T$ are
the fields with vanishing VEV's. Then these mixing terms do not contribute to
the Higgs masses because of $\langle S\rangle =
\langle T\rangle =0$ (if we add the soft breaking
terms, $\langle S,T \rangle $ will be of
${\cal O}(M_{weak})$, so the contribution
to the Higgs masses will be also suppressed by a factor of $M_{weak}/M_{GUT}$
which is exactly what we need). Thus such fields with vanishing VEV's can yield
the desired suppression of the mixing terms.

One such an example could be a superpotential of the form

\begin{equation}
\label{sol}
aS(\bar{H}H-\alpha M^2)+bT({\rm Tr}\Sigma^3-\beta N^3).
\end{equation}
(We take a cubic term in $\Sigma$
because the trace of $\Sigma$ vanishes and a quadratic term would give
SU(35) accidental symmetry. $N$ and $M$ should be singlets with respect to
the SU(6) group so that their VEV doesn't break the symmetry further and also
to avoid a larger accidental symmetry.)

The equation of motion for the $S,T$ singlets sets
\begin{eqnarray}
\label{eom}
\langle \bar{H}H \rangle = \alpha \langle M^2 \rangle \nonumber \\
{\rm Tr} \langle \Sigma^3 \rangle =\beta \langle N^3 \rangle ,
\end{eqnarray}
while the $S$ and $T$ VEV's vanish because of the other equations of motion.

The problem with this model is that the VEV's of the fields $M,N$ and
consequently of $\Sigma ,H,\bar{H}$ are not determined.
To find an acceptable theory based on the superpotential of eq. \ref{sol}
we need to reintroduce the GUT-scale into
our theory by setting the $M$ and $N$ VEV's to the GUT-scale by hand.
 The origin of this new
scale in the theory could be for example a condensation scale of a strongly
interacting gauge group (other than the SU(6)). We assume that for some
reason the fields
$M,N$ acquire VEV's of ${\cal O} (M_{GUT})$. Then these VEV's can be
communicated to the fields $\Sigma ,H,\bar{H}$ without introducing mixing
terms through the eqs. \ref{eom}.
(In other words we could say that an effective tadpole term in the
superpotential for the fields $S$ and $T$ is generated by integrating
out heavy fields that have VEV's of the order of the GUT scale which
spontaneously break the discrete symmetry.)
But even if we set the $M,N$ VEV's
to the desired value the $H,\bar{H},\Sigma $ VEV's are still not
totally determined. This is done by the D-terms and the soft breaking terms.
The D-terms vanish if $\langle H \rangle = \langle \bar{H} \rangle$ and
$\langle \Sigma \rangle$ is diagonal.  Now adding the soft breaking
terms will shift the values of the VEV's by terms of the order of the
weak scale and also lifts the very high degeneracy of the $\Sigma$ vacua.
 Eq. \ref{eom} fixes only ${\rm Tr} \Sigma^3$.
After we add the soft breaking terms the only possible $\Sigma$  vacua are
those which break SU(6) to  SU(n)$\otimes$SU(6-n)$\otimes$U(1) depending
on the values of the parameters of the soft breaking terms.

To forbid the direct mixing terms (those without the fields $S,T$)
we should set the discrete  charges of $\Sigma , H\bar{H}$ to be small,
so that the mixing terms require high powers of these fields.
If the discrete symmetry is not an R-type then by choosing the charges
of  $\Sigma , H\bar{H}$ the charges of the other fields are
already determined.
 For example if we take
the charges as $Q_S=\frac{15}{16}$, $Q_{\bar{H}H}=Q_{M^2}=\frac{1}{16}$,
$Q_T=\frac{20}{21}$ and $Q_{\Sigma^3}=Q_{N^3}=\frac{1}{21}$, the first
mixing term without $S,T$ is $(\bar{H}H)^{16}\Sigma^{63}$, or
$\bar{H}H,\Sigma $ exchanged to $M^2$ or $N$, while the mixing terms
involving $S,T$ are automatically suppressed by a factor
$(\frac{M_{weak}}{M_{GUT}})$.
(These mixing terms are just the products of the operators in the two sectors.)

We can go further and forbid even the mixing terms that include $S,T$ by
promoting the discrete symmetry to an R-symmetry.
For example if we assign the R-charge for the fields $Q_S=\frac{1}{55}$,
$Q_{\bar{H}H}=\frac{2}{165}=Q_{N^2}$, $Q_T=\frac{1}{44}$,
$Q_{\Sigma} =Q_{M}=\frac{1}{396}$, and that of the superpotential is
$\frac{1}{33}$ then all mixing is forbidden to more than 50 orders.

\subsection{Summary}
In this section we have presented three different type of models that all
yield acceptable theories. We have shown how to circumvent the
difficulties of section 3 and build natural theories with sufficiently
suppressed mixing terms.

One might think that the above arguments for building a superpotential are
true only for the SU(6) model we have
considered. In the next section we show that alternative
models based on the idea of two noninteracting sectors and
either an SU(n), SO(n), or E$_6$ gauge group
which do not have additional light doublet or triplet fields
are trivial generalizations of the model we have considered,
and therefore yield no more compelling solutions. (The only restrictive
requirement we will make for this proof
is that during SSB the unbroken subgroups are only in
regular embeddings of the full group).

\mysection{Models with Larger Gauge Symmetry}

In this section we discuss the possibilities of generalizing the SU(6)
model based on the groups SU(n), SO(n) and E$_6$ (these include all groups that
are capable of admitting complex representations).
We restrict our search to models that use regular group embeddings only
and have no light triplets which could result in too large proton decay.

Let us first discuss the criteria that  a realistic model where the Higgses
are pseudo-Goldstone bosons have to fulfill. We are looking for
grand unified models with gauge group $G$ that have an accidental symmetry
of the Higgs part of the superpotential. This accidental symmetry
is a consequence of the existence of two sectors (A and B)
that are not mixed with
each other; thus the global symmetry is $G\otimes G$.
We do not consider the other kind of models
 when representations pair up to representations of a bigger
group since
these models are  necessarily fine tuned; a larger numbers of sectors
is overly cumbersome and we do not consider it.
Throughout this chapter (and Appendix A) we consider only models where the
unbroken subgroups of the full symmetry group are in regular embeddings
(it is very difficult to derive general results if we also allow the use
of special embeddings).
The fields in sectors A and B develop VEV's such that the gauge group $G$
breaks to the standard model
group SU(3)$\otimes $SU(2)$\otimes $U(1). We are interested
in all possible symmetry breaking patterns for $G=$ SU(n), SO(n) and E$_6$
for which the resulting uneaten
PGB's  fulfill the following four requirements.

1. At least two SU(2) doublets are uneaten PGB's of the global symmetry which
can be identified with the Higgs fields of the MSSM.

2. The extra uneaten PGB's besides the two doublets (if there are some)
do not destroy the successful prediction of unification of coupling
constants. We know that generally adding extra light particles to the MSSM
spectrum destroys the prediction for $\sin^2\theta_W$. The only case when this
does not happen is when  we add full SU(5) multiplets. Then the one loop value
of $\sin^2\theta_W$ is unchanged.

3. The extra uneaten PGB's do not give rise to too fast proton decay. This
can be prevented if the extra light fields do not contain SU(3) triplets
whose quantum number are equal to those of
$d^c$ ($d$ denotes the down quark), because
this is the field that if light generically mediates proton decay.
Although generally the presence of such
light triplets spoils
proton decay, we can avoid this problem by forbidding the coupling of these
new triplets to ordinary matter (the phenomenological implications
of such new triplets were given in \cite{Bar1}). Nevertheless we find it
 natural to consider only models that do not contain extra $d^c$'s
(that is they do not contain for example extra $5$'s of SU(5)).

4. The extra uneaten PGB's do not destroy asymptotic freedom of QCD. For
example if the extra particles are only in SU(3) triplets than their number
is constrained to be less than six. Thus for example  the extra
uneaten PGB's can not be combined into $10+\bar{10}$ of SU(5) because
this contains exactly 6 triplets.

Summarizing the requirements on the uneaten PGB's for a realistic model we find
that in addition to the 2 doublets that are identified with the Higgses of the
MSSM we can have only full SU(5) multiplets that do not contain
triplets with the quantum numbers of $d^c$ and the number of extra triplets
is constrained to be less than six.

Let us now discuss how many uneaten PGB's we get from the SSB
in the case of $G=$ SU(n). The general method  was presented in \cite{Bar1}.
The accidental global symmetry is SU(n)$\otimes $SU(n).
We have two sectors of the superpotential that are not mixed (that's how
we get the accidental global SU(n)$\otimes$SU(n) symmetry). We denote
these sectors A and B. The fields in sectors A and B develop
VEV's such that the gauged SU(n) breaks to SU(3)$\otimes $SU(2)$\otimes $U(1).
We are interested in all possible SSB patterns which fulfill the above
four requirements of a realistic model and the additional requirement
of the use of regular embeddings only.

The VEV's of the fields in
one of the two sectors (for example A) have to split the SU(3)$_c$ from
the SU(2)$_w$ which are embedded into the SU(n) gauge group. (In the SU(6)
model sector A is identified with the sector containing the adjoint
field. The adjoint in the SU(6) model breaks SU(6) to
SU(4)$\otimes$SU(2)$\otimes$U(1) and thus splits the SU(3)$_c$ subgroup
from the SU(2)$_w$.)

If we consider only regular embeddings of subgroups then the most general
subgroups of SU(n) are products of SU(k) factors and U(1) factors. Thus
the fields of
sector A break the global  SU(n) to
\beq
{\rm SU}({\rm{m}_1)}\otimes
{\rm SU}({\rm m}_2)\otimes {\rm SU}({\rm m}_3) \otimes \ldots
\otimes {\rm possible}\;
{\rm U}(1)\; {\rm  factors}
\eeq
where we can assume that SU(2)$_{weak} \subset $SU(m$_1$),
SU(3)$_{c}\subset$ SU(m$_2$). So the VEV's of the fields of
this  sector split the SU(3)$_c$ subgroup
from the SU(2)$_w$ subgroup.
Because of our choice of embeddings for the SU(2)$_w$ and SU(3)$_c$ we
know that
 ${\rm m}_1 \geq 2$, ${\rm m}_2\geq 3$ and also
 $\sum_{i=1}^l {\rm m}_i \leq
{\rm n}$ must be true.

The combination of the VEV's of the
fields of sectors A and B must break the gauge group to
SU(3)$\otimes $SU(2)$\otimes $U(1), this means that the gauged
subgroup SU(m$_1$)$\subset $SU(n) breaks to
SU(2),
the SU(m$_2$)$\subset $SU(n) breaks to SU(3), and all other subgroups must be
totally broken.

Let us first concentrate on the PGB's we get from the SU(m$_1+$m$_2$) subgroup
of SU(n) (this is where the standard model group is also embedded). According
to our assumptions, the
fields of sector A break this SU(m$_1+$m$_2$)
subgroup  to SU(m$_1$)$\otimes$SU(m$_2$)( possibly $\otimes$U(1)), and the
combination of the VEV's of both sectors
must give SU(3)$\otimes$SU(2)$\otimes$U(1) as the unbroken
gauge group. This can be achieved only if the fields of sector B break
the SU(m$_1+$m$_2$) subgroup either to SU(5) or to
SU(3)$\otimes$SU(2)$\otimes$U(1).
First we calculate how many uneaten PGB's we get from this SU(m$_1+$m$_2$)
subgroup if the fields of
sector B break this SU(m$_1+$m$_2$) subgroup
 to SU(5). Uneaten PGB's are states in the SU(n) adjoint corresponding to
 generators that
are broken by the fields of both
sectors. It is easy to check that from the breaking
 SU(m$_1$+m$_2$)$\otimes $SU(m$_1$+m$_2$)$\rightarrow$
SU(m$_1$)$\otimes$SU(m$_2$)$\otimes$SU(5) we get the following uneaten PGB's:
\begin{eqnarray}
\label{GB1}
({\rm m}_1-3) [(1,2)_{\frac{1}{2}} +(1,2)_{-\frac{1}{2}}]
+({\rm m}_2-2)[(3,1)_{-\frac{2}{3}}+( \bar{3},1)_{\frac{2}{3}}]
+{\rm singlets},
\end{eqnarray}
where the PGB's are denoted according to
their SU(3)$\otimes$SU(2)$\otimes$U(1)
transformation properties. (If fields of sector B break SU(m$_1+$m$_2$)
 to SU(3)$\otimes $SU(2)$\otimes$U(1) instead
of SU(5) we get extra uneaten PGB's in the representations
$(3,2)+(\bar{3},2)$ of SU(3)$\otimes $SU(2).
These representations
 contain 6 doublets, so even if we combine them with some other
 light particles to full SU(5)
representations (to preserve the prediction for $\sin^2\theta_W$)
we will have at least 6 extra triplets added to the
MSSM particle content. This is unacceptable
because of the asymptotic freedom of QCD. Thus the fields of
sector B must break the  SU(m$_1+$m$_2$) subgroup to SU(5).)
 To avoid the presence of triplets in eq. \ref{GB1}
(these triplets are contained in $5$ and $\bar{5}$'s of SU(5) so they
generally give rise to fast proton decay) we
need ${\rm m}_2=2$. Because from the states outside the SU(m$_1$+m$_2$)
subgroup we can get only full SU(5) representations as uneaten PGB's
we  need to set ${\rm m}_1=4$ to get the two light uneaten doublets.
Thus ${\rm m}_1=4, {\rm m}_2=2$, and the fields of sector A break this
SU(m$_1+$m$_2$) subgroup (which is an SU(6) because m$_1=2$, m$_2=4$)
to SU(4)$\otimes $SU(2)$\otimes $U(1) and the fields of sector B to SU(5).

Now we are able to discuss the SSB of the full global
SU(n)$\otimes$SU(n).
According to our assumptions and the previous argument,
the fields of  sector A break SU(n) to
SU(4)$\otimes$SU(2) $\otimes$SU(m$_3$)$\otimes \ldots \otimes$ U(1) factors.
This means that the generators in the two off-diagonal $5$ by $ ({\rm n}-6)$
matrices of the SU(n) adjoint
are already broken by the fields of sector A, and all these states transform
nontrivially under SU(3)$\otimes $SU(2)$\otimes$U(1) of the MSSM.
If we suppose that the fields of sector B
break SU(n) to a smaller subgroup than SU(n$-1$) then some
of these states (in the offdiagonal 5 by n$-6$  matrices of the adjoint)
were also uneaten
PGB's and we would get additional light doublets and
triplets. These states transform all according to
 $5$ or $\bar{5}$ of SU(5). Thus they would give
rise to proton decay.
 This means that the only acceptable breaking for the fields of sector B
is to break SU(n) to SU(n$-1$). But because the unbroken gauge group is
SU(3)$\otimes $SU(2)$\otimes $U(1) and we get this by combining the VEV's
of both sectors
we must have ${\rm m}_i=0$ for $i>2$ in sector A. Otherwise
we would get a bigger unbroken gauge group than
SU(3)$\otimes$SU(2)$\otimes$U(1). This means that for sector A,
SU(n)$\rightarrow$SU(4)$\otimes$SU(2)$\otimes$U(1), while for sector B,
SU(n)$\rightarrow$SU(n$-1$).

 Thus we have
shown that if we rely only
 on regular embeddings of subgroups into bigger groups
  the only acceptable pattern of symmetry breaking in SU(n) models is
SU(n)$\otimes $
 SU(n)$\rightarrow $SU(4)$\otimes $SU(2)$\otimes $U(1)$\otimes $SU(n$-1$).
This generates two light doublets and additional light
SU(3)$\otimes $SU(2)$\otimes $U(1) singlets.
This is an obvious generalization of the SU(6)
model, and we have also proven that
there is no other way in SU(n) models using only regular embeddings
to implement the idea of having
the Higgses as PGB's. Using the same requirements we
 show in  Appendix A that there is no satisfactory
model based on the gauge groups SO(n) or E$_6$.
Thus we see that all realistic models that have the Higgses as PGB's
and which do not require special embeddings are
based on the unification group SU(n), and for every n there is
only one possibility for the pattern of symmetry breaking.

Finally we give the necessary fields for the generalized SU(n) theories
with ${\rm n}\geq 6$,
where
the symmetry breaking pattern of the accidental global symmetry is
SU(n)$\otimes $SU(n)$\rightarrow $
SU(4) $\otimes $SU(2)$\otimes $U(1)$\otimes $SU(n$-1$). This symmetry breaking
can be most naturally achieved with the following fields:
an adjoint ($\Sigma$) and ${\rm n}-6$ pairs
of vectors ($H_i,\bar{H}_i$) in
sector A of the superpotential and another pair
of vectors ($H',\bar{H}'$) in  sector B with the following VEV's:

\begin{eqnarray}
\langle \Sigma \rangle = V \left( \begin{array}{cccccccc} a & & & & & & & \\
 & a & & & & & & \\ & & a & & & & & \\ & & & \ddots & & & & \\ & & & & a & & &
 \\ & & & & & b & & \\ & & & & & & \ddots & \\ & & & & & & & b \end{array}
\right) \begin{array}{c} \\ \\ k \\ \\ \\ \\ m \\ \end{array}
\end{eqnarray}

\begin{eqnarray}
\langle H_i \rangle = \langle \bar{H}_i \rangle = V_i \left( \begin{array}{c}
0 \\ 0 \\ \vdots \\ 1 \\ \vdots \end{array} \right)
\end{eqnarray}

\begin{eqnarray}
\langle H'\rangle = \langle \bar{H}' \rangle =W  \left( \begin{array}{c}
1\\ 0\\ 0 \\ \vdots \\ 0 \end{array} \right)
\end{eqnarray}
with $ka+mb=0$, $W> V,V_i$
(in order to get the correct order of gauge symmetry breaking
SU(n)$\rightarrow$SU(n$-1$)$\rightarrow$SU(3)$\otimes$SU(2)$\otimes$U(1) ),
 the smallest of $V,V_i$ must be the GUT
scale, and the ones in the VEV's of $V_i$ are organized such that together
with the adjoint they break SU(n) to SU(4)$\otimes $SU(2)$\otimes $U(1).

The minimal anomaly free fermion content of an SU(n) theory \cite{King}
is one two index antisymmetric representation (denoted by $T^{ab}$) and
${\rm n}-4$ conjugates of the defining
 representation ($t_a$) per family. If we only want
to have the usual light SU(5) fermions $( 10+\bar{5})_i$,
then the remaining SU(5) nonsinglet fields must be heavy. The
SU(5) content of $T^{ab}$ and $t_a$ is
\begin{eqnarray}
\label{dec}
&& T^{ab}=  10+({\rm n}-5) 5 + {\rm singlets} \nonumber \\
&& t_a = \bar{5} + {\rm singlets}
\end{eqnarray}
This means that in the case of 3 families we must have $3({\rm n}-5)$
heavy $5$'s
and $\bar{5}$'s.

 Operators that could give GUT-scale masses to  fields of eq. \ref{dec}
are:
\begin{eqnarray}
&& \lambda_{ij}T^{ab}_{(i)}t_{a(j)}\bar{H}_b  \nonumber \\
&& \lambda_{ij}^m T^{ab}_{(i)}t_{a(j)}\bar{H}_b'^{m}
\end{eqnarray}
where $i,j$ denote generation indices while $a,b$ denote SU(n) indices.
The first term yields 3 heavy $ 5$'s and $ \bar{5}$'s
while the second term
yields $3({\rm n}-6)$ heavy $ 5$'s and $ \bar{5}$'s.

Thus it is possible to split the heavy fields from the light in a natural
way for every n. The bigger problem is to account for a heavy
(${\cal O} (M_W)$) top quark. The method of \cite{Bar3} (see
also Appendix B) does not work for
${\rm n}>6$. In this case
the smallest self adjoint representation that does not
have a heavy mass term is a representation with ${\rm n}/2$ antisymmetrized
indices for n even but n/2 odd.
 But the SU(5) content of this field (called $M$) is

\begin{eqnarray}
M=\left( \begin{array}{c} n-5 \\ \frac{n}{2}-1 \end{array} \right) (5+
 \bar{5})+
\left( \begin{array}{c} n-5 \\ \frac{n}{2}-2 \end{array} \right)
(10+ \bar{10}) + {\rm singlets}.
\end{eqnarray}
out of which we only want to have one of the $10$'s light,
 and all other fields
must be heavy (and also one of the $10$'s
of $T^{ab}$ must then be heavy).
We show that with these fields we can not have a consistent theory.
The reason is that with only these fields and assuming matter
parity we can have only one renormalizable operator containing $M$ in the
superpotential, $M\Sigma M$. But this does not give GUT-scale mass to any
of the fields in $M$ because of the antisymmetry of the indices. (In the
case of SU(6) there was another operator $20 H 15$ that made the extra
unwanted fields heavy, but in the  case n$ >6$  the operator $M H_i T$ is not
SU(n) invariant). Thus the masses for these new fields coming from $M$ can
arise only from the nonrenormalizable operators. But this means that all
the fields from $M$
have masses less than the GUT-scale. Because $M$ is a very
big representation of SU(n) it contains a lot of new triplets. (The lowest
possible value of n is 10, because this is the first case when
n is even and n/2
is odd. If n=10, the extra number of triplets is 70.) This badly destroys
asymptotic freedom and is therefore not acceptable (in the case of n=10
the couplings blow up before we reach $M_{GUT}$, and the ${\rm n}>10$
cases are even worse).
Also because we don't have renormalizable mass terms for the fields in $M$
the masses of the triplets will be lower than $M_{GUT}$ which is dangerous
for proton decay.

It is even more difficult to find an acceptable superpotential for the n
$>6$ models than in the n=6 case. The extra difficulty comes from the
sector where we need to combine an adjoint and n$-$6 pairs of vectors to get
the required symmetry breaking. In this sector we have to mix the adjoint
and the vectors to avoid a bigger accidental symmetry than SU(n)$\otimes
$SU(n). Thus we necessarily have terms of the form $\bar{H}_i \Sigma H_j$.
But the presence of such terms generally requires $\langle H \rangle =0$.
We illustrate this for the a superpotential with SU(6) fields:
\begin{equation}
\label{ill}
W=\frac{1}{2}M {\rm Tr}\Sigma^2+\frac{1}{3} \lambda {\rm Tr}\Sigma^3 +\alpha
\bar{H}\Sigma H .
\end{equation}
The equation of motion for $\Sigma$ will be
\begin{equation}
\label{sigeom}
M\Sigma_{ij} +\lambda (\Sigma^2_{ij}-\frac{1}{6}{\rm Tr}\Sigma^2 \delta_{ij}) +
\alpha \bar{H}_iH_j =0.
\end{equation}
If we assume that
\begin{eqnarray}
\label{VEVs}
\langle \Sigma \rangle = V \left( \begin{array}{cccccc} 1 & & & & & \\ &
1 & & & & \\ & & 1 & & & \\ & & & 1 & & \\ & & & & -2 & \\ & & & & & -2
\end{array} \right) , \langle H \rangle =\langle \bar{H} \rangle =W
\left( \begin{array}{c} 1 \\ 0 \\ 0 \\ 0 \\ 0 \\ 0 \end{array} \right),
\end{eqnarray}
we get for the $\Sigma $ equation of motion that
\begin{eqnarray}
\label{incon}
(MV+\lambda V^2) \left( \begin{array}{cccccc} 1 & & & & & \\ &
1 & & & & \\ & & 1 & & & \\ & & & 1 & & \\ & & & & -2 & \\ & & & & & -2
\end{array} \right) +\alpha W^2 \left( \begin{array}{cccccc} 1 & & & & & \\ &
0 & & & & \\ & & 0 & & & \\ & & & 0 & & \\ & & & & 0 & \\ & & & & & 0
\end{array} \right)=0,
\end{eqnarray}
which in turn requires that $W=0$. Thus it is difficult to find a
superpotential where adjoints and vectors are mixed and both have
comparable VEV's. (We need to introduce extra fields that mix with both the
adjoint and the vectors, but which  have zero VEV's while they allow for
nonzero VEV's of both the adjoint and the vectors.)

\mysection{Conclusions}

The biggest problem of the SU(6) model where the Higgses are PGB's of
the spontaneously broken accidental SU(6)$\otimes $SU(6) symmetry is
to construct a superpotential with some possible discrete symmetries
that yields naturally the correct VEV's for the fields and mixing is
forbidden to sufficiently high order.
We have shown how to construct such models.
 We have also shown why we expect these to be the
only type of models which will work.
 The reason for this is  that without having either a
small mass scale or fields with zero VEV
in our theory we generally need either fine
tuning of the parameters or have the symmetry breaking mixing terms
not sufficiently suppressed.
We have demonstrated that we can naturally make use of the mass scales
present in a theory to build an acceptable model.
We have also shown that we can build models without using small mass
scales by using fields that have naturally zero VEV's.
We presented three possible models with sufficiently suppressed mixing terms.
In the first example we made use of the supersymmetry
breaking scale that is necessarily present in every theory.
In the second model we did not use small mass scales but had singlets with
zero VEV's.
In the
third model we assumed the dynamical appearance of the GUT scale.

We have also shown that there are only very few realistic models that
use only regular group embeddings and have
the Higgses as PGB's. All of them are trivial generalizations of the SU(6)
model built on the gauge group SU(n).

We conclude that it is possible to successfully implement the
Higgs as Goldstone boson scheme in  models
where only discrete global  symmetries or
gauge symmetries  are assumed to be exact to all orders.
Based on our considerations we expect that the simplest models
successfully implementing the idea of having the Higgses as
pseudo-Goldstone particles will
have features of those presented in section 4.

\section*{Acknowledgements}

C.C. and L.R. are grateful to Greg Anderson for useful discussions and
especially for pointing out the fine tuning in ref. \cite{Bar2}.
Z.B. thanks  Oleg Kancheli for useful discussions. Z.B. and L.R.
thank S. Pokorsky and the other organizers for their hospitality at the
Warsaw SUSY Meeting 'Physics from Planck Scale to Electroweak Scale',
where part of this work was done.

\section*{Appendix A: SO(n) and E$_6$}
\setcounter{equation}{0}
\renewcommand{\theequation}{A.\arabic{equation}}

In this appendix we show that it is not possible to generalize the SU(6)
model to orthogonal groups or to E$_6$  if one allows only for regular
embeddings.

If the gauge group is SO(n) then  the SU(2)$_w$ must be embedded into an
SO(4) subgroup and the SU(3)$_c$ into an SO(6) subgroup of SO(n), thus
\begin{eqnarray}
{\rm SU}(3)\otimes {\rm SU}(2) \subset {\rm SO}(6)\otimes
{\rm SO}(4) \subset {\rm SO}(10).
\end{eqnarray}
We know that SO(6)$\sim $SU(4), SO(4) $\sim$ SU(2)$\otimes $SU(2), so the
SU(3)$\otimes $SU(2) transformation properties of the states in the
adjoint ($45$) of the SO(10)
subgroup of SO(n) where the standard model group is embedded are
\begin{eqnarray}
&& [( 8,1)+(3,1)+(\bar{3},1)+(1,1)] + \nonumber \\ && [(1,3)+
3(1,1)]+ 2 [(3,2)+(\bar{3},2)] +{\rm singlets}.
\end{eqnarray}

Now we can apply the same method we used for SU(n).
We have again two sectors, A and B.
The VEV's of the fields of
one of the sectors (A) have to split the SU(2)$_w$ and SU(3)$_c$ subgroups,
and thus SO(n) breaks either to
SO(m$_1$)$\otimes $SO(m$_2$)$\otimes \ldots $
or to SU(n$_1$)$\otimes$SU(n$_2$)$\otimes \ldots$.
In the first case (orthogonal subgroups)
we have the embedding
SU(3) $\subset $ SO(m$_1$), SU(2) $\subset $ SO(m$_2$). The combination
of the VEV's  of the
two sectors
 must break the gauged SO(m$_1$) to SU(3) and the gauged SO(m$_2$) to SU(2).
If the fields of
the sector A  break the SO(m$_1$+m$_2$) subgroup
 to SO(m$_1$)$\otimes $SO(m$_2$),
  then we need the fields of sector B
to break the SO(m$_1+$m$_2$) subgroup to
SU(5) so  that the combination  of the VEV's of the two
sectors break the gauged  SO(m$_1$+m$_2$)  to
SU(3)$\otimes $SU(2)$\otimes$U(1).
(Should we choose SU(3)$\otimes $SU(2)$\otimes$U(1) instead of SU(5)
for the breaking of SO(m$_1+$m$_2$) by the fields of sector B  we would
get more PGB's just like in the case of SU(n) models). One can show that
from this SO(m$_1$+m$_2$) subgroup alone we get the following PGB's:
\begin{eqnarray}
({\rm m}_1-6)2(1,2) + ({\rm m}_2-4)[(3,1)+(\bar{3},1)]+
(3,2)+(\bar{3},2) ,
\end{eqnarray}
where the PGB's  are
denoted according to their SU(3)$\otimes$SU(2) transformation properties.
So no matter what ${\rm m}_1$ and ${\rm m}_2$ are we get the light
$(3,2)+(\bar{3},2)$
particles  which are unacceptable because they destroy unification (if we
want them to combine with some other
light states  to full SU(5) representations, we get at least 6
triplets because we already have 6 doublets. Thus in this case asymptotic
freedom of QCD would be destroyed).

The other possibility is that SU(3)$\otimes $SU(2) is embedded in unitary
groups in sector A,
SU(3)$\otimes $SU(2) $\subset $SU(m$_1$)$\otimes $SU(m$_2$) $\subset $
SO(2m$_1$)$\otimes $SO(2m$_2$)$\subset $SO(2m$_1+$2m$_2$).
In this case we need to break this SO(2m$_1+$2m$_2$) subgroup by the
fields of
sector B to an SO(10) subgroup in order to break the gauged
  SO(2m$_1$+2m$_2$) subgroup in the
right way. The resulting uneaten PGB's from this SO(2m$_1$+2m$_2$)
subgroup can be shown to be (according to their SU(3)$\otimes$SU(2)
transformation properties):
\begin{eqnarray}
& & (2{\rm m}_2+{\rm m}_1-7)[(3,1)+(\bar{3},1)]+
\nonumber \\ & & 2(2{\rm m}_1+{\rm m}_2-8) (1,2) +
{\rm singlets}
\end{eqnarray}
Because ${\rm m}_1\geq 3,\; {\rm m}_2 \geq 2$, we can
get zero triplets (they are again part of $5$'s
and $\bar{5}$'s of SU(5) so their presence would generally give
rise to proton decay) only if ${\rm m}_1=3$ and
${\rm m}_2=2$.
This just means that SO(2m$_1$+2m$_2$)=SO(10) and the fields of
sector A break SO(10) to SU(3)$\otimes $SU(2)$\otimes$U(1), and the
fields of sector B do not break this SO(10) subgroup of SO(n).
 But subsequently from the other subgroups
we can get only full SU(5) multiplets,
so we do not have doublet PGB's. This means that there
is no SO(n) theory with only
regular embeddings that would successfully generalize the SU(6) model.
(If the SU(3)$\otimes $SU(2) are mixed in a unitary
and an orthogonal
subgroup in sector A that is  SU(3)$\subset$SO(m$_1$) and
SU(2)$\subset$SU(m$_2$) or SU(3)$\subset$SU(m$_1$) and SU(2)$\subset$SO(m$_2$)
we get the same analysis as for the case when SU(3)$\subset$SO(m$_1$)
 and SU(2)$\subset$SO(m$_2$) but with even
more PGB's, so these cases do not work either.)

In the case of the E$_6$ group we can have 3 possible regular embeddings of
SU(3)$\otimes $ SU(2)$\otimes $U(1) into E$_6$. These cases correspond
to the following maximal subgroups of E$_6$: SU(6)$\otimes $SU(2),
SO(10)$\otimes $U(1) or SU(3)$\otimes $SU(3)$\otimes $SU(3).

If we place SU(3)$\otimes $SU(2) into the SU(6) subgroup we can
repeat our argument  that in one sector the global SU(6) must break to
SU(4)$\otimes $SU(2)$\otimes $U(1)
while in the other sector to SU(5). The adjoint
of E$_6$ is $78$, its decomposition under SU(6)$\otimes $SU(2) is
\begin{equation}
78= ( 35,1)+(1,3)+(20,2).
\end{equation}
Because fields of both both sectors
break the adjoint along the SU(6)$\otimes$SU(2)
subgroup the states in
$(20,2)$ must be uneaten PGB's. Their SU(5)
 decomposition is $2(10+\bar{10})$, and these would be light. But
this would yield 12 new light triplets that destroy asymptotic freedom of
QCD.
We also saw that based on orthogonal groups we can not build an
acceptable model, so the SO(10)$\otimes $U(1) case is excluded as well.

The last possibility for E$_6$ models is to embed
SU(3)$\otimes $SU(2)$\otimes $U(1)
into the SU(3)$\otimes $ SU(3)$\otimes $SU(3) subgroup. The
decomposition of $ 78$ under this subgroup is
\begin{equation}
78=(8,1,1)+(1,8,1)+(1,1,8)+(3,3,\bar{3})+
(\bar{3},\bar{3},3) .
\end{equation}
The fields in one of the sectors (A) need to break E$_6$ at least to
SU(3)$\otimes $SU(3)$\otimes $SU(3) to split the SU(3)$_c$ from SU(2)$_w$.
Because of the unification of couplings this must happen
at the GUT scale, and the fields
of the other sector (B) must break E$_6$ at a scale above
$M_{GUT}$ to either SU(5), SU(6) or SO(10). (Otherwise we get
large threshold corrections for the RG value of $\sin^2\theta_W$.)
If the fields of sector B break E$_6$ to SU(5)
 than only particles in $(3,2)+(\bar{3},2)$ will be eaten from
sector A which still leaves at least fields transforming according to
SU(3)$\otimes$SU(2) as
\begin{equation}
 3[(\bar{3},1)+(3,1)]+2[(\bar{3},2)+(3,2)]
\end{equation}
uneaten. (For the
unification of couplings it would be sufficient that fields of sector B
break the global E$_6$ to SU(5)$\otimes$possible other factor like
SU(5)$\otimes$SU(2), but
these possible other factors do not play any role since their
presence can only change the number of SU(5) singlets. Thus the same
analysis would apply. The same is true for the next two cases when
the fields of sector B break E$_6$ to SO(10) or SU(6).)

If the fields of sector B break E$_6$
to SO(10) than one can show that the uneaten PGB's are at least
(according to their SU(3)$\otimes $SU(2) transformation properties)
\begin{equation}
(3,1)+(\bar{3},1)+2[(\bar{3},2)+(3,2)].
\end{equation}
Finally if the fields of sector B break E$_6$ to SU(6) the uneaten
nontrivially transforming PGB's are
at least
\begin{equation}
 2[(\bar{3},1)+(3,1)]+3[(\bar{3},2)+(3,2)].
\end{equation}
This means that in neither case are we able to get the desired light
particle spectrum (in each case we
get more than six triplets), so there
is no way to have a realistic model built on E$_6$ that has
the Higgs particles as PGB's using these embeddings.

\section*{Appendix B: Fermion Masses}
\setcounter{equation}{0}
\renewcommand{\theequation}{B.\arabic{equation}}

In this appendix we show that it is possible to extend the successful
picture of fermion masses of refs. \cite{Bar3,Bar4} with the discrete
charges in accordance with a realistic Higgs sector.
For the Higgs sector we will use the superpotential presented in section 4,
extending the discrete symmetry to the fermion sector.
We will see a model that is consistent both in the
Higgs sector and in the fermion sector. This serves as an existence proof
for realistic models implementing the idea of having the Higgs fields as
pseudo-Goldstone bosons.

First we briefly review the model of refs. \cite{Bar3,Bar4} for fermion
masses in the SU(6) model.
The minimal anomaly free fermion content of SU(6) that includes one
generation of light fermions is
\begin{equation}
15+\bar{6}+\bar{6}',
\end{equation}
where $15$ is the two index antisymmetric representation and $\bar{6}$ is
the conjugate of the defining representation. One can add any self adjoint
representation and maintain anomaly cancellation. Generally  self adjoint
representations have invariant mass terms so it is no use adding
them to the fermion content. But there are some special cases when this mass
term vanishes. For example if we add just one representation $20$ of SU(6)
(three index antisymmetric representation) then the mass term for this
vanishes by antisymmetry (in general if we have odd number of $20$'s one
of them will have a vanishing mass). We remark that the addition of a $20$
to the usual particle content of the theory destroys asymptotic freedom
of the SU(6) gauge coupling. But this is not a problem since with only
one $20$ the increase of the coupling is very slow, its value increases
only a few percent between the GUT and the Planck scale.

The idea of \cite{Bar3} is to add the extra $20$ to the fermion content
which will be then
\begin{equation}
\label{ferm}
(15+\bar{6}+\bar{6}')_i+20, ~~~~ i=1,2,3
\end{equation}
The SU(5) decomposition of these fields is
\begin{eqnarray}
& & 20=10+\bar{10}, \nonumber \\
& & 15=10+5, \nonumber \\
& & \bar{6}=\bar{5}+1.
\end{eqnarray}
Then the renormalizable Yukawa couplings have the form
\begin{equation}
\label{yuk}
\lambda^{(1)} 20\Sigma 20 +\lambda^{(2)} 20\; H\; 15_i + \lambda_{ij}^{(3)}
15_i\bar{H}\bar{6}'_j, ~~~~ i,j=1,2,3
\end{equation}
($i,j$ denote generation indices). If we insert the VEV's of
$H,\bar{H},\Sigma$ and the Higgs doublets into $\Sigma$ (if
$\langle H\rangle \gg \langle \Sigma \rangle$, the Higgs doublets live
almost entirely in the $\Sigma$ field, see eqs. \ref{higgs},\ref{higgs2})
we get the following mass terms:
\begin{equation}
\label{mass}
\lambda^{(2)}\langle H \rangle 10_i \bar{10}+\lambda_{ij}^{(3)} \langle H
\rangle 5_i\bar{5}'_j +\lambda^{(1)} Q\;u^c\;h_2, ~~~~i,j=1,2,3,
\end{equation}
where the decomposition of $10$ of SU(5) is $Q+u^c+e^c$. The fermion
fields in (\ref{ferm}) contain altogether four $10$'s, six $\bar{5}$'s,
three $5$'s and one $\bar{10}$ of SU(5). From (\ref{mass}) we see that
out of these fields one combination of $10$'s, three of $\bar{5}$'s and
the three $5$'s and the $\bar{10}$ will get masses of ${\cal O}(M_{GUT})$,
so the light fermion spectrum is the desired
\begin{equation}
3\times(10+\bar{5}),
\end{equation}
while only one light fermion (namely, the up type quark contained in 20)
gets a mass from the renormalizable interaction with the Higgs doublet.
The reason is that the couplings of the 20-plet explicitly
violate the global SU(6)$_\Sigma\otimes$SU(6)$_H$ symmetry, so that
the Higgs doublet $h_2$ has  non-vanishing coupling to the
up type quark from 20, which can be identified with the top quark.
Thus, the top mass is naturally in the 100 GeV range.
Other fermions stay massless at
this level, unless we invoke the higher order operators
explicitly violating the accidental global symmetry.

To go further we need to introduce nonrenormalizable operators to give
masses to the other fermions. Generally, these operators explicitly
violate the accidental global symmetry, since they include both the
$\Sigma$ and $H,\bar H$ fields, so that they can provide nonvanishing
Yukawa couplings to the Higgs doublets, though suppressed by $M_{Pl}$.
In refs. \cite{Bar3,Bar4} these operators
were obtained from heavy fermion exchange
\cite{Frogatt}. For this purpose a specific set of heavy
(Planck-scale) vectorlike fermion superfields in nontrivial SU(6)
representations was introduced whose couplings with
the light fermions and the Higgs superfields yielded the needed
structure of nonrenormalizable operators
(together with flavor-blind discrete $Z_2$ \cite{Bar3} or
$Z_3$ \cite{Bar4} symmetries to forbid some unwanted operators).
For example, in ref. \cite{Bar3} the
relevant nonrenormalizable operators coming from the specified heavy
fermion superfield exchanges were specified by
\begin{eqnarray}
& & \frac{1}{M_{Pl}} (20\Sigma )H\; 15_i, \; i=1,2,3, \nonumber \\
& & \frac{1}{M_{Pl}^2} 20 (\bar{H}\Sigma \bar{H})\bar{6}_3 \nonumber \\
& & \frac{1}{M_{Pl}^2} 15_i (\Sigma \bar{H})(\Sigma \bar{6}_j), i,j=2,3 .
\end{eqnarray}
The first operator gives mass to the $c$ quark, the second to the $b$
and $\tau$, and the third to the $s$ and $\mu$. These masses will have
a proper hierarchy provided that $\langle H \rangle >> \langle \Sigma
\rangle$. In the model of ref.\cite{Bar3} the first generation fermions were
left massless, however in the model of ref. \cite{Bar4} they can also get
masses of the needed value.

However in our approach all nonrenormalizable operators that are not
forbidden by some symmetry are present in the superpotential.
In other words, we would like to obtain all masses in general operator
analysis, not relying on heavy fermion exchange mechanism \cite{Frogatt}
with specified fields. Therefore a `flavor democratic' approach to
fermion masses (which means that there are no `family symmetries'
that would distinguish among the generations) is out of question: it
would yield too heavy first generation masses.
Thus we will need to use family symmetries in constructing the fermion
mass terms. The simplest way is just to extend the discrete symmetries
used for the stabile picture in the Higgs sector also to the
fermion sector.

For the demonstration, we will use the Model 1 presented in Section 4,
which is based on a $Z_3\otimes Z_n^{(1)}\otimes Z_n^{(2)}$ discrete
symmetry. The discrete charge assignments are given in table 2.
\begin{table}
\label{tab}
\[ \begin{array}{|c|c|c|c|c|c|c|c|c|c|c|c|} \hline
  &  H       &    \bar H   & \Sigma_1     & \Sigma_2    &     20      &
    15_3      &   15_{2,1}  &\bar{6}_{3,1} & \bar{6}_2   & \bar{6}_3'  &
 \bar{6}_{2,1}' \\ \hline Z_4^{(1)}
& \frac{1}{n} &    0        &    0        &     0       &     0       &
 -\frac{1}{n} &-\frac{1}{n} &    0        & \frac{1}{n} & \frac{1}{n} &
\frac{1}{n}  \\ \hline  Z_4^{(2)}
&    0        & \frac{1}{n} &    0        &     0       &     0       &
     0        &     0       &-\frac{2}{n} &-\frac{1}{n} & -\frac{1}{n}&
 -\frac{1}{n}
\\ \hline  Z_3
&    0        &     0       & \frac{1}{3} &-\frac{1}{3} & \frac{1}{3} &
 -\frac{1}{3} &     0       &    0        &     0       & \frac{1}{3} &
     0
\\ \hline \end{array} \]
 \caption{The charge assignments of the chiral superfields under the discrete
 $Z_n^{(1)}\otimes Z_n^{(2)}\otimes Z_3$ symmetry.}
 \end{table}
The Higgs sector is given by the superpotential of eq. \ref{goodsup}.
 Because we have now two adjoint
fields in one of the sectors of the accidental symmetry the $h_1$, $h_2$
Higgs doublet fields live in a linear combination of them. One can show
that the uneaten PGB doublets are given by
\begin{eqnarray}
\label{Higgs}
& & h_1= \cos\alpha(\cos\gamma\,h_{\Sigma_1} +
\sin\gamma\,h_{\Sigma_2}) - \sin\alpha\, h_H  \nonumber \\
& & h_2= \cos\alpha(\cos\gamma\,\bar{h}_{\Sigma_1} +
\sin\gamma\,\bar{h}_{\Sigma_2}) - \sin\alpha\, \bar{h}_{\bar{H}}
\end{eqnarray}
where $\tan\gamma=V_2/V_1$ and $\tan\alpha=3V/U$. Here $\langle H \rangle
=\langle \bar H \rangle = U$, $\langle \Sigma_{1,2} \rangle=V_{1,2}$,
and $V=(V_1^2+V_2^2)^{1/2}$.

If $V_1 \sim V_2 \ll U$, as it occurs e.g. for $n=6$, then the Higgs
doublets are dominantly contained in $\Sigma_1$ and $\Sigma_2$ while almost
not contained in $H$ and $\bar{H}$.

The allowed Yukawa couplings in this
model together with their physical role are listed below.

- $15_i \bar{H}\bar{6}'_i$: makes the extra $\bar{5}$'s and 5's heavy.

- $20 H  15_3 $: makes the extra 10 and $\bar{10}$ heavy.

- $ 20 \Sigma_1 20$: yields heavy top quark.

- $ \frac{1}{M^2} 20 \Sigma_2 \bar{H}\bar{H} \bar{6}_3$: defines
$\bar{6}_3$ state and yields b and $\tau$ masses.

- $\frac{1}{M^2} 20 \Sigma_2 H 15_2$: defines 15$_2$ state and yields
 charm mass via c-t mixing.

- $ \frac{1}{M^2} 15_{2,1} \Sigma_1\Sigma_2 \bar{H}\bar{6}_2$: gives s,
$\mu$ masses and Cabbibo mixing.

- $\frac{1}{M^4} 15_{2,1} \Sigma_1\Sigma_2 \bar{H} (\bar{H}H)
\bar{6}_{3,1}$: gives d,e masses and 1-3 mixing.

- $ \frac{1}{M^2} 15_{2,1} H \Sigma_1 \Sigma_2 H 15_{2,1} $: gives u mass.

M denotes the suppression scale of the nonrenormalizable operators.
We denote $\epsilon_H=\frac{\langle H \rangle}{M}$, $\epsilon = \frac{\langle
\Sigma \rangle}{\langle H \rangle}$. As we noted before, for $n=6$
$\langle H \rangle$ is at an intermediate scale scale between $\langle
\Sigma \rangle$ and $M_{Pl}$. Since $\langle \Sigma \rangle=M_{GUT}
\simeq 10^{16}\,$GeV is fixed (as the $SU(5)$ scale) by the gauge
coupling unification, we obtain
$\epsilon_H \sim \epsilon \sim 1/30$, which can explain the
fermion mass hierarchy. A somewhat better fit can be obtained with the
slightly larger value $\epsilon_H \sim 0.1$. This could occur if the above
listed operators are generated by heavy fermion exchange (the
heavy particles should be below the Planck scale, with masses
$M\sim 10^{18}\,$GeV). The desired value of the
parameter $\epsilon $ should remain $\sim 1/30$ which fits  perfectly to
the light generation fermion masses.

 The physical consequences of the
above listed operators can be summarized as follows ($\lambda$ denotes the
Yukawa couplings of the MSSM, while $\theta_{ij}$ denotes the mixing
angle of the i'th and j'th generation):

- $\lambda_t \sim 1 $

- $\lambda_{b,\tau}\sim \epsilon_H^2$, $\lambda_b=\lambda_{\tau}$.

- $\lambda_c\sim \epsilon_H^2$

- $\sin \theta_{23} =\sqrt{\frac{\lambda_c}{\lambda_t}}\sim \epsilon_H$.

- $ \lambda_{s,\mu}\sim \epsilon \epsilon_H^2$, but the ratio
$\lambda_s/\lambda_{\mu}$ is not fixed (because there is more than one
operator due to the different possible
contractions of indices in the operator $15_{2,1} \Sigma_1\Sigma_2
\bar{H} \bar{6}_2$).

- $\lambda_{d,e}\sim \epsilon \epsilon_H^4$, but the ratio of the
couplings is not fixed again. The ratio $\lambda_e/\lambda_{\mu}$ is
of order $\epsilon_H^2 \sim \lambda_c/\lambda_t$. However
$\lambda_d$ is a somewhat small.

- $\lambda_u \sim \epsilon\epsilon_H^3$. The ratio
$\lambda_u/\lambda_c$ is of order $\epsilon\epsilon_H \sim
\lambda_\mu/\lambda_\tau\sqrt{\lambda_c/\lambda_t}$.

- $\sin \theta_{12}\sim {\cal O}(1)$ (Cabbibo angle).

- $ \sin \theta_{13}\sim \frac{\lambda_d}{\lambda_b}$.

All these consequences (except the down mass which must be enhanced
by introducing a large Clebsch coefficient)
are in qualitative agreement with the experimental values, provided that
$\tan\beta$ is small (close to 1). In fact, here
we used the general operator analysis consistent with the gauge
$SU(6)$ and discrete symmetries. By addressing the specific heavy
fermion exchanges, one could also fix the relative Clebsch factors
between the down quark and charged lepton masses \cite{Bar3,Bar4}.
Thus we have shown that a consistent model based on the SU(6) gauge
group and discrete symmetries can be constructed. In this model the
accidental global SU(6)$\otimes$SU(6) symmetry is preserved by
nonrenormalizable operators in the Higgs superpotential up to sufficiently
high order terms, so that the Higgs doublets are PGB's without
any fine tuning. On the other hand,
Yukawa terms explicitly violating the SU(6)$\otimes $SU(6)
symmetry yield the necessary Yukawa couplings for the light fermion masses.

\end{document}